\pdfoutput=1
\documentclass[letterpaper]{ar-1col}

\usepackage{aas_macros}
\usepackage{natbib}%

\def\lapp{\ifmmode\stackrel{<}{_{\sim}}\else$\stackrel{<}{_{\sim}}$\fi}
\def\gapp{\ifmmode\stackrel{>}{_{\sim}}\else$\stackrel{>}{_{\sim}}$\fi}

\jname{Xxxx. Xxx. Xxx. Xxx.}
\jvol{AA}
\jyear{YYYY}
\doi{10.1146/((please add article doi))}

\begin{document}

% Page header
\markboth{Kaspi \& Beloborodov}{Magnetars}

% Title
\title{Magnetars}

%Authors, affiliations address.
\author{Victoria M. Kaspi,$^1$ Andrei M. Beloborodov$^2$
\affil{$^1$Department of Physics and McGill Space Institute, McGill University, Montreal, Canada H3W 2C4; email: vkaspi@physics.mcgill.ca}
\affil{$^2$Physics Department and Columbia Astrophysics Laboratory, Columbia University, New York, USA 10027; email:  amb@phys.columbia.edu}
}

% First page note
\firstpagenote{First page note to print below DOI/copyright line.}

%Abstract
\begin{abstract}
Magnetars are young and highly magnetized neutron stars which display
a wide array of X-ray activity including short bursts, large outbursts, giant flares
and quasi-periodic oscillations, often coupled with interesting timing behavior
including enhanced spin-down, glitches and anti-glitches. The bulk of this
activity is explained by the evolution and decay of an ultrastrong magnetic field,
stressing and breaking the neutron star crust, which in turn drives twists of
the external magnetosphere and powerful magnetospheric currents.
The population of detected magnetars has grown to about 30 objects and shows
unambiguous phenomenological connection with very highly magnetized radio pulsars.
Recent progress in magnetar theory includes explanation of the hard X-ray component
in the magnetar spectrum and development of surface heating models, explaining
the sources' remarkable radiative output.
\end{abstract}

%Keywords, etc.
\begin{keywords}
Neutron stars, magnetic fields, radio pulsars, flares, X-ray astronomy, X-ray outbursts.
\end{keywords}
\maketitle

%Table of Contents
\tableofcontents

% Text Box
%\begin{textbox}
%\section{TEXT BOX HEAD}
%Text box text. Text box text. Text box text. Text box text. Text box text. Text box text.
%\subsection{Text Box Sub-head}
%Text box text. Text box text. Text box text. Text box text. Text box text. Text box text.
%\subsubsection{Text Box Subsub-head}
%Text box text. Text box text. Text box text. Text box text. Text box text. Text box text.
%\end{textbox}
%
%\begin{textbox}
%Text box text. Text box text. Text box text. Text box text. Text box text. Text box text.
%Text box text. Text box text. Text box text. Text box text. Text box text. Text box text.
%Text box text. Text box text. Text box text. Text box text. Text box text. Text box text.
%\end{textbox}

\section{INTRODUCTION}
\label{sec:intro}

Magnetars are a class of young neutron stars that exhibit
dramatic variability across the electromagnetic spectrum -- particularly
at X-ray and soft $\gamma$-ray energies -- ranging from few millisecond
bursts to major month-long outbursts.  Some magnetar outbursts include X-ray
and soft $\gamma$-ray flares
that briefly outshine the entire cosmic hard X-ray sky put together.  
Magnetar emission is powered by the decay of enormous internal magnetic fields, 
which is why \citet{dt92a} coined the name `magnetar.'
%Though their
%neutron-star nature is, at this point, impossible to escape, that these
%sources manage such hyper-Eddington-luminosity events is one of their
%most astonishing properties and one which points to the origin of their
%whimsical name:  these are neutron stars powered by the decay of the largest magnetic fields
%known in the Universe.  
%In contrast to conventional radio pulsars as well as to
%accreting binary neutron stars 
%(such as those
%previously reviewed by \cite{tm77} or \cite{ts86},  
%as well as discussed by \"Ozel in
%this volume), 
%magnetars must harness a novel power source, not only to explain
%their `giant flares,' but also in order to understand their more commonly observed phenomenology.
%John Wheeler reportedly expressed surprise upon the discovery of radio pulsars that
%neutron stars come equipped with a ''handle and a bell\footnote{This quote was reported by
%\cite{mt77} but its exact origin was not specified.};" this was presumably in reference
%to their surprising radiative properties, notably regular bright radio pulsations.
%Were Wheeler to describe magnetars, his list of equipment would have been much longer and
%would have included far more volatile devices.
%\footnote{See the online magnetar
%catalog at www.physics.mcgill.ca/$\sim$pulsar/magnetar/main.html as well as \cite{ok14}.}  
The known magnetar population today consists of just 29 sources
however they are likely to represent at least 10\% of the young neutron star population
(and possibly a much larger fraction).  There are nearly 
1000 papers on the subject, a testament 
%This is a testament not only to the interesting puzzle magnetars have represented since their discovery
%in the late 1970's, but also 
to their relevance to many branches of astrophysics today,
from gravitational waves 
%\citep[e.g.][]{dgps15} 
to superluminous supernovae 
%\citep[e.g.][]{wwdw15} 
to gamma-ray bursts 
%\citep[e.g.][]{gmk+15} 
to Fast Radio Bursts.
%\citep[e.g.][]{tkp16}.
 
%of the 884 refereed
%papers containing the word 'magnetar,' at the time of this writing, 89 were published in 2015, and
%72 thus far in 2016.
%These recent magnetar publications involve astrophysics ranging from gravitational waves \citep[e.g.][]{dgps15} to
%superluminous supernovae \citep[e.g.][]{wwdw15,bsb15} to gamma-ray bursts \citep[e.g.][]{gmk+15} to Fast
%Radio Bursts \citep[e.g.][]{pc15,tkp16}.
%Although all interesting topics, space limitations preclude all but this brief mention of magnetars'
%importance beyond just their own physics and place in the neutron star zoo.

%In this review, after a historical description of the path that has led
%to our current understanding of magnetars and their relationship to other neutron stars,
%we provide an overview of the presently known magnetar population.
%Then we examine in more depth hallmark magnetar phenomena.
%, including the plethora
%of temporal behaviors seen, both in the rotational evolution of these objects as
%well as in their radiative output.  The latter is examined more specifically  
%in \S\ref{sec:spectral} where their spectra are considered.  
%The second half of this review then concerns itself with our present theoretical understanding of the physics
%of magnetars.
%, with an effort to connect back to their observable properties described
%first.  
%We %will conclude with our view of what are the most important outstanding
%issues in this field.
Several other magnetar reviews have been written, some primarily
observational \citep{re11}, some primarily theoretical \citep{tzw15}, and some, like this one,
a combination \citep{wt06, mpm15}.  For reviews putting magnetars into the context of the broader neutron-star
population, see \citet{kas10} and \citet{kk15}.

\subsection{Historical Overview}
\label{sec:history}

Historically, magnetars first appeared in astronomy under names ``Soft Gamma Repeaters (SGRs)" and 
``Anomalous X-ray Pulsars (AXPs)."
The first publication reporting a magnetar detection
was in 1979 when repeated bursts were seen by space-based hard
X-ray/soft gamma-ray instruments aboard the interplanetary space probes
Venera 11 and 12 \citep{mgg79,mgi+79,mg81}.
Although first thought to have the same origin as the classical gamma-ray
bursts (GRBs), repeated bursts, including one enormous flare on 1979
March 5 from the direction of the star-forming Dorado region in the
Large Magellanic Cloud \citep[LMC][]{mgi+79} rendered this unlikely.
%to remark, ``This event differs sharply in its properties from all the
%$\gamma$-ray bursts we have detected previously, and it is of outstanding
%interest.''  
The repeated bursts had decidedly softer spectra than those of
the gamma-ray bursts, hence the designation as ``soft gamma repeaters''
(SGRs).  Additional soft gamma-ray bursts from what is known today
to be magnetar SGR 1900+14 \citep{mgg79,mg81} 
provided clear evidence of a new class of Galactic high-energy sources. 

The neutron-star nature of the SGRs was clear early on.
The 8-s pulsations seen in the
declining flux tail following the flare from the LMC source, subsequently
known as SGR 0526$-$66, were 
strongly suggestive of a neutron-star origin \citep{mgi+79}, a conclusion supported by
the coincidence of the pulsar with the supernova remnant N49 \citep{cdt+82}.
The 8-s period was, however, notably longer than that of other young neutron
stars like the 33-ms Crab pulsar, and initially an unsteadily accreting neutron star in a binary
was thought to be the best explanation for the bursts. 

It was not until several years later that the distinct class was fully recognized,
when SGRs 0526$-$66 and 1900+14 were joined by a
a third Galactic source, SGR 1806$-$20, which exhibited roughly 100 bursts between 1978 and 1986, with
most in 1983 \citep{lfk+87,knc+87}. 

%\subsection{The Magnetar Model}

The magnetar model was born in considering
the possibility of amplification of a seed helical magnetic field under dynamo action
in a proto neutron star immediately following a core-collapse supernova \citep{dt92a}.
%The authors coined the term `magnetars,' and 
\citet{td95} and \citet{td96a} demonstrated that SGR phenomena are nicely
explained by spontaneous magnetic field decay
serving as an energy source for the transient bursts and outbursts as well as
for the persistent emission seen in these sources. 
Their rationale hinged on both neutron-star rotational dynamics and energetics
arguments:  given the location of the 8-s pulsar SGR 0526$-$66 in the LMC supernova remnant
N49 (and later noting the
central 7-s pulsar in supernova remnant CTB 109)
a surface dipolar field strength of order $10^{14}-10^{15}$~G is required to brake
the pulsar from few-ms birth periods in $\sim 10^4$~yr, the typical lifetime of a
supernova remnant \citep[see also][]{pac92}.  
Further, such a field, particularly if stronger inside the star, could
provide a large energy reservoir to explain SGR activity.
%Theoretical arguments based on required field strength
%for spontaneous decay to occur \citep{gr92} also supported the field estimates, as
%did the requirement for magnetic confinement of plasma presumably emitted suddenly at the start
%of a giant flare, whose tail lasted several minutes \citep{td95}.  
Thus, multiple
lines of reasoning argued for the existence of magnetic field strengths several orders
of magnitude larger than had been previously estimated for any radio pulsar or accreting
X-ray pulsar.  An unambiguous prediction was made:  SGR periods must be increasing with time,
since so large a field must brake the star on $\sim 10^3 - 10^4$-yr time scales.

%\begin{textbox}
%The dipolar component of the surface magnetic field of a rotation-powered pulsar is
%commonly estimated from the expression $B_s = 3.2 \times 10^{19} \sqrt{ P \dot{P} }$~G,
%derived under the assumption of magnetic braking due to dipole radiation in a vacuum.
%Although radio pulsar magnetospheres are almost certainly filled with plasma, 
%magnetohydronamical simulations \citep{spi06} have shown that the estimate is
%valid up to a factor of 2--3.  Although the energy source for radiation in magnetars
%is magnetic field decay rather than rotational energy in
%radio pulsars, the dipole braking formula is expected
%to be roughly valid for magnetars as well.
%\end{textbox}

In 1998, the first measurement of an SGR spin-down rate was reported \citep{kds+98}, and both
its sign and magnitude (in this case for SGR 1806$-$20) provided stunning confirmation 
of the magnetar model predictions:
under the standard assumption of a magnetic dipole braking, in which the surface
dipolar magnetic field strength $B$ is estimated from $B = 3.2 \times 10^{19} \sqrt{P \dot{P}}$~G,
a field strength of $\sim 8 \times 10^{14}$~G was inferred.
This measurement was followed shortly thereafter by an analogous one for SGR 1900+14 \citep{ksh+99}.
%, and
%eventually a likely similar spin-down measurement for SGR 0526$-$66 \citep{kkm+03}.
The direct measurement of the model-predicted spin-down is what sealed the identification of
SGRs as magnetars for most of the astrophysics community.

%\subsection{Anomalous X-ray Pulsars}
%\label{sec:axps}

Meanwhile, a puzzle seemingly unrelated to SGRs was developing.
\citet{gf80}, using the {\it Einstein} observatory, reported ``an extraordinary
new celestial X-ray source," a supernova remnant, CTB 109, with a bright X-ray source
at the center.  \citet{fg81} reported that this
source exhibits strong pulsations with a period of 3.5 s (later realized to be the
second harmonic of a 7-s fundamental).
%, though \citet{fg83}
%later realized this was the second harmonic and the fundamental
%is 7~s \citep{fg83}).  
The discovery of two additional such few-second X-ray pulsars, albeit not in supernova remnants,
1E~1048.1$-$5937 \citep{scs86} and later
4U 0142+61 \citep{hel94,ims94}, suggested a new source class, as did their 
unusually soft X-ray spectra.  However
such long pulse periods and the unusual spectra were generally interpreted to indicate a 
new type of low-mass X-ray binary, with the pulsations accretion-powered \citep[e.g.][]{ms95,smi96}.
The name ``anomalous X-ray pulsar'' (AXP) was introduced by \citet{vtv95}.

\citet[][hereafter TD96]{td96a} 
made a crucial suggestion that AXPs may be related to SGRs. Although surprising at the time, from today's
hindsight it seems 
obvious:  if AXPs were young neutron stars with a puzzlingly strong
energy source and few-second periods, they could well be magnetars as well.
%More detailed considerations of AXP properties, including ages, luminosities, and spectra,
%as well as of the viability of different models -- including accreting low-mass binaries
%and merged white dwarfs, led TD96 to suggest that AXPs are another form of
%magnetar, less active than SGRs, but still radiating X-rays powered by a
%decaying magnetic field.  
TD96 remarked, regarding AXPs, ``And, in
the future, one might expect to detect SGR burst activity
from one or more of these objects!''

This expectation was unambiguously confirmed six years later by the discovery of SGR-like bursts from two AXPs
\citep{gkw02,kgw+03}.
As described later in this review, bursting is today known to be a characteristic property of AXPs,
so much so that the line between them and SGRs, for all intents and purposes, no longer exists.

Another key expectation of the magnetar model was that these objects would be
prolific glitchers (TD96).  Glitches are sudden spin-ups of the neutron star
that are commonly observed in young radio pulsars like the Crab pulsar.
%Glitching in magnetars was argued by TD96 to be a natural consequence of sudden crustal
%fractures due to magnetic-field stresses and was argued to be likely to be accompanied
%by changes in X-ray luminosity.  
The discovery of the first magnetar glitch by
\citet{klc00} and the subsequent realization that such events are ubiquitous in
these sources and are often -- though not always -- accompanied by X-ray outbursts
\citep[e.g.][]{dkg08}
is another verified magnetar model prediction. 

\section{OVERVIEW OF THE KNOWN MAGNETAR POPULATION}
\label{sec:overview}

\subsection{Basic Properties}
\label{sec:basic}

Recently, the first magnetar catalog was published \citep[][hereafter OK14]{ok14}, and includes a detailed compendium of the properties
of the known magnetars.   Here we summarize these properties and refer the reader for details to that
paper or, for the most recent updates, to the online catalog\footnote{www.physics.mcgill.ca/$\sim$pulsar/magnetar/main.html}.

The vast majority of known magnetars were discovered via their short X-ray bursts, thanks to sensitive all-sky
monitors like the Burst Alert Telescope (BAT) aboard {\it Swift} and the Gamma-ray Burst Monitor (GBM) aboard {\it Fermi}.  These instruments
were designed to study gamma-ray bursts so are fine-tuned to finding brief, bright bursts over the full sky.  
Thus there is strong
bias in the known magnetar population toward sources most likely to burst.  
That nearly all known magnetars share common
spin properties and high spin-inferred surface dipolar magnetic fields is a powerful statement regarding which objects
in the neutron-star population are burst-prone.

In short, apart from the hallmark X-ray activity that defines the class (see \S\ref{sec:bursts}),
magnetars are observed to produce X-ray pulsations in the period range 2--12 s (ignoring two
faster-rotating sources that only briefly exhibited magnetar-like properties; 
see Table~\ref{ta:srcs} and \S\ref{sec:highB}).  Magnetars
are, without exception, spinning down, with spin-down rates that imply spin-down time scales ($\sim P/\dot{P}$)
of a few thousand years, suggesting great youth.  The spin-down luminosity $\dot{E} \equiv 4 \pi^2 I \dot{P}/P^3$, where
$I \simeq 10^{45}$~g~cm$^2$ is the stellar moment of inertia, is usually far smaller than the persistent quiescent
X-ray luminosity of the sources (see Table~\ref{ta:srcs}).
Moreover, these spin-down rates imply, for 20/23 of the sources
for which it has been measured, $B>5\times 10^{13}$~G, with the vast majority over $10^{14}$~G.  
Whereas most radio pulsars are thought to be born with periods
of at most a few hundred milliseconds, that the shortest known {\it bona fide} magnetar
has a relatively long 2-s period in spite of a young age
is surely a result of rapid magnetic braking. 
%inevitable in so highly magnetized a neutron star.
%Faster rotators may exist, but they slow down so rapidly (in under $\sim$1000 years), they are rare to observe.
The long period cutoff of 12~s has been more of a puzzle, which is related to the life-time
of magnetar activity \citep[e.g.][]{cgp00,vrp+13}.
The small observed
ranges in $P$ and $B$ are contrasted by a far larger range in quiescent X-ray luminosity, spanning 
$\sim 10^{30}$~erg~s$^{-1}$ up to $2 \times 10^{35}$~erg~s$^{-1}$ in the 2-10-keV band.  
In fact, the distribution of quiescent luminosities appears somewhat bi-modal, with the
brighter group the `persistent' magnetars and the fainter ones the `transient' magnetars.  The latter 
show greater dynamic range in their outbursts.
This large range in luminosity is presently an interesting puzzle, as is
the small range in period (see \S\ref{sec:pulsations}).
%The X-ray pulse profiles in all cases are broad, with duty cycles of well over 50\%, and significant energy dependence of the pulse morphology.
%Pulse fractions range from a few percent to over 70\%.
In the soft X-ray band, magnetar spectra are fairly well described by a blackbody and in some
cases an additional power-law component, while in the hard X-ray band significant spectral hardening occurs such that
in some cases the energy spectrum {\it rises}, at least as far as has been detected (typically until $\sim$60~keV).   
%Magnetar spectra are described in more detail in \S\ref{sec:spectral} and discussed in \S\ref{sec:andrei}.
Magnetar emission has been seen in some cases at radio, infrared and optical wavelengths.
%The multi-wavelength properties of magnetars are discussed further
%in \S\ref{sec:multitemporal}, \S\ref{sec:spectra_oir} and \S\ref{sec:spectra_radio}.

One source is notable and not included in Table~\ref{ta:srcs} --
the central source of the supernova remnant RCW~103, 1E~161348$-$5055.  It
shows a strange 6.67-hr X-ray periodicity with a variable pulse profile, as well as repeated large X-ray outbursts
\citep{dcm+06b}.  \citet{aeb+16} and \citet{rbe+16} reported on the discovery of a bright magnetar-like burst from the
source, coupled with another large X-ray flux outburst.  The source thus bears all the hallmarks of a magnetar,
except for the bizarrely long spin period, which cannot be from simple magnetic braking.  
%which is 2000 times longer than the previous record holder!  
%Simple magnetic braking cannot accomplish this for any reasonable initial set of parameters given the young age of the supernova remnant.
The long period may be explainable with a fall-back disk 
that slows down the initially faster-rotating neutron star 
\citep{li07,ha17}.  
%but leaves open the question of why such a mechanism occured only in this source
%and not in the any of the other known magnetars.  
%In this magnetar picture, the source should at least today
%be spinning down due to conventional magnetic braking.  However, such slow down will be very difficult to measure:
%for a $10^{15}$-G magnetar spinning once every 6.67 hr, the expected $\dot{P}$ is predicted to result in a

% Example of a Table
\begin{table}
\tabcolsep3.5pt
\caption{Known Magnetars and Magnetar Candidates$^{\rm a}$}
\label{tab1}
\begin{center}
\begin{tabular}{@{}l|c|c|c|c|c|c|c@{}}
\hline
Name$^{\rm b}$ & $P$ & B$^{\rm c}$ & Age$^{\rm d}$ & $\dot{E}$$^{\rm e}$ & D$^{\rm f}$ & $L_X$$^{\rm g}$ & Band$^{\rm h}$ \\
     & (s) & ($10^{14}$ G) & (kyr) & $10^{33}$ erg~s$^{-1}$ & (kpc) & $10^{33}$ erg~s$^{-1}$ &  \\
\hline
CXOU J010043.1−-721134 & 8.02 & 3.9 & 6.8 & 1.4 & 62.4 & 65 & ... \\
4U 0142+61 & 8.69 & 1.3 & 68 & 0.12&  3.6 & 105 &  OIR/H \\
SGR 0418+5729 & 9.08 & 0.06 & 36000 & 0.00021 & $\sim$2 & 0.00096 &  ... \\
SGR 0501+4516 & 5.76 & 1.9 & 15 & 1.2& $\sim$2 & 0.81 & OIR/H  \\
{\bf SGR 0526$-$66} & 8.05 & 5.6 & 3.4 & 2.9 & 53.6 & 189 &  ... \\
1E 1048.1−-5937 & 6.46 & 3.9 & 4.5 & 3.3 & 9.0 & 49 &  OIR \\
(PSR J1119$-$6127) & 0.41 & 4.1 & 1.6 & 2300 & 8.4 & 0.2 & R/H \\
1E 1547.0−-5408 & 2.07 & 3.2 & 0.69 & 210 & 4.5 & 1.3 & O?/R/H \\
PSR J1622--4950 & 4.33 & 2.7 & 4.0 & 8.3 & $\sim$9 & 0.4 & R \\
SGR 1627$-$41 & 2.59 & 2.2 & 2.2 & 43 & 11 & 3.6 & ...\\
CXOU J164710.2--455216 & 10.6 & $<$0.66 & $>$420& $<$0.013 & 3.9 & 0.45 &  ...  \\
1RXS J170849.0--400910 & 11.01 & 4.7 & 9.0 & 0.58 & 3.8 & 42 & O?/H \\
CXOU J171405.7--381031 & 3.82 & 5.0 & 0.95 & 45 & $\sim$13 & 56 & ...\\
SGR J1745--2900 & 3.76 & 2.3 & 4.3 & 10& 8.3 & $<$0.11 &  R/H \\
{\bf SGR 1806$-$20} & 7.55 & 20 & 0.24 &45 & 8.7 & 163 &  OIR/H \\
XTE J1810--197 & 5.54 & 2.1 & 11 & 1.8& 3.5 & 0.043 & OIR/R \\
Swift J1822.3--1606 & 8.44 & 0.14 & 6300 & 0.0014 & 1.6 & $>$0.0004 &  ... \\
SGR 1833--0832 & 7.56 & 1.6 & 34 & 0.32 & ... & ... & ... \\
Swift J1834.9--0846 & 2.48 & 1.4 & 4.9 & 21 & 4.2 & $<$0.0084 & ... \\
1E 1841--045 & 11.79 & 7.0 & 4.6 & 0.99& 8.5 & 184 & ...  \\
(PSR J1846$-$0258) & 0.327 & 0.49 & 0.73 & 8100 & 6.0 & 19 & ... \\
3XMM J185246.6+003317 & 11.56 & $<0.41$ & $>1300$ & $<0.0036$& $\sim$7 & $<0.006$ & ... \\
{\bf SGR 1900+14} & 5.20 & 7.0 & 0.9 & 26& 12.5 & 90 & H \\
SGR 1935+2154 & 3.24 & 2.2 & 3.6 & 17 & ... & ... & ...\\
1E 2259+586 & 6.98 & 0.59 & 230 & 0.056 & 3.2 & 17 &  OIR/H\\
{\it SGR 0755$-$2933} & ... & ... & ...& ... & ... & ... & ... \\
{\it SGR 1801$-$23}   & ... & ... & ... & ... & ... & ... & ...\\
{\it SGR 1808$-$20}    & ... & ... & ... & ... & ... & ... & ...\\
{\it AX J1818.8$-$1559}     & ... & ... & ... & ... & ... & ... & ...\\
{\it AX J1845.0$-$0258} & 6.97 & ... & ... & ...& ... & 2.9 &  ... \\
{\it SGR 2013+34 } & ... & ... & ... & ... & ... & ... & ...\\
\hline
\end{tabular}
\end{center}
\begin{tabnote}
$^{\rm a)}$All tabulated values from OK14;
$^{\rm b)}$Sources in {\bf bold} have had giant flares. Sources in {\it italics} are candidates only. Sources in parentheses are
normally rotation-powered pulsars;
$^{\rm c)}$Spin-inferred magnetic field strength; 
$^{\rm d)}$Characteristic age from $P/2\dot{P}$; 
$^{\rm e)}$Spin-down luminosity;
$^{\rm f)}$Distance; 
$^{\rm g)}$Unabsorbed quiescent luminosity in the 2--10-keV band for the distance provided; 
$^{\rm h)}$OIR=Optical/Infrared counterpart. R=Radio counterpart. H=Hard X-rays detected.
\end{tabnote}
\label{ta:srcs}
\end{table}

% Name, P, B, tau, dist, Lx, Association, OIR?, radio?

\subsection{Spatial Distribution}
\label{sec:spatial}

One of the best-determined aspects of magnetars'  spatial distribution in the Galaxy is their strict confinement
to the Galactic Plane.  As shown by OK14, the scale height of known magnetars is only 20--30~pc, in spite
of the vast majority of known objects having been discovered through their X-ray bursts by 
spatially unbiased all-sky X-ray monitors.  
%This scale height reflects uncertainties in distance (see below) and
%accounts for small biases in discovery.  
This scale height is far smaller than that of the radio pulsar
population, clearly indicating great youth in magnetars.  A
200 km~s$^{-1}$ spatial velocity will have moved an object by $\sim$20~pc in $10^5$~yr.  
Direct proper motion
measurements for magnetars have found a
weighted average 200~km~s$^{-1}$ with standard deviation 100~km~s$^{-1}$
\citep{tck13}, somewhat lower than that of the radio pulsar population \citep[e.g.][]{acc02,bfg+03,fk06}.
Thus magnetars typically cannot be much older than $10^5$~yr, and are generally much younger.  
%Clearly models involving white dwarf progenitors are strongly at odds with this basic property.

%Conventional radio pulsars generally have relatively large spatial velocities \citep[e.g.]{cc98,acc02,bfg+03,fk06}
%likely owing to asymmetric natal kicks.
%, but 
%early papers on the subject suggested magnetars should have even higher space velocities.  
%Motivated by the significant
%offset of  SGR 0526$-$66 from the center of its putatively associated
%supernova remnant N49 \citep{rkl94} -- though this same offset has also been used to argue against
%an association; \citep{gsgv01} -- \citet{dt92a} proposed multiple mechanisms for the production
%of kick velocities as high as 1000~km~s$^{-1}$ in magnetars.   
%Measurements of magnetar proper motions
%combined with best available distance estimates yield the following 6 tangential spatial velocities for magnetars:
%212$\pm$35~km~s$^{-1}$ for 1E 1810$-$197 \citep{hcb+07}, 
%280$\pm$120~km~s$^{-1}$ for 1E 1547.0$-$5408 \citep{dcrh12}, 
%130$\pm$30~km~s$^{-1}$ for SGR 1900+14 and 350$\pm$100~km~s$^{-1}$ for SGR 1806$-$20
%\citep{tck12},
%157$\pm$17~km~s$^{-1}$ for 1E~2259+586 and 102$\pm$26~km~s$^{-1}$ for 4U~0142+61
%\citep{tck13}.  The weighted average is 

\begin{figure}
\begin{minipage}{2.9in}
\includegraphics[scale=0.45]{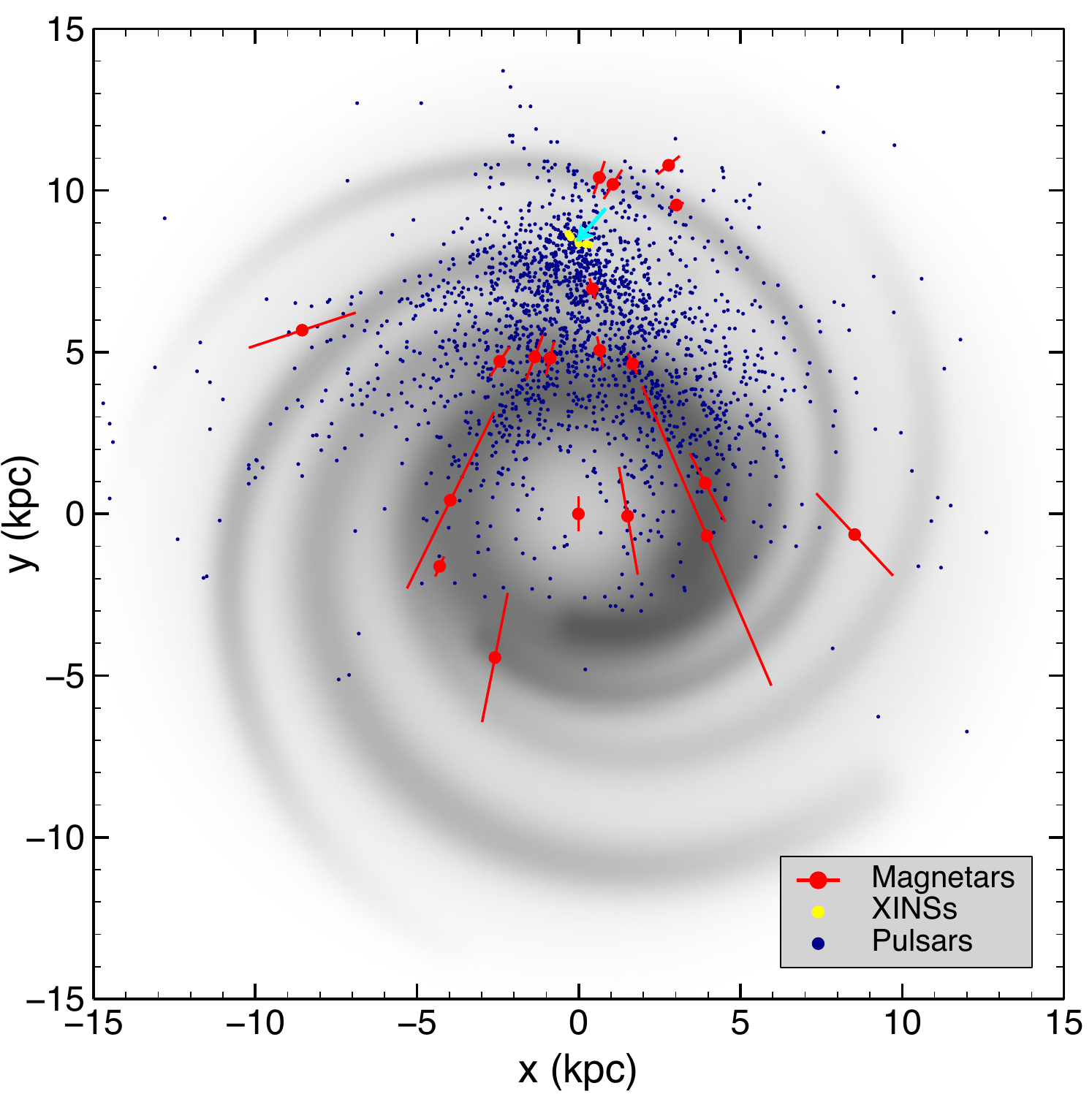}
\end{minipage}
%\hfill
\hspace{-1.55in}
\begin{minipage}{2.9in}
\includegraphics[scale=0.4]{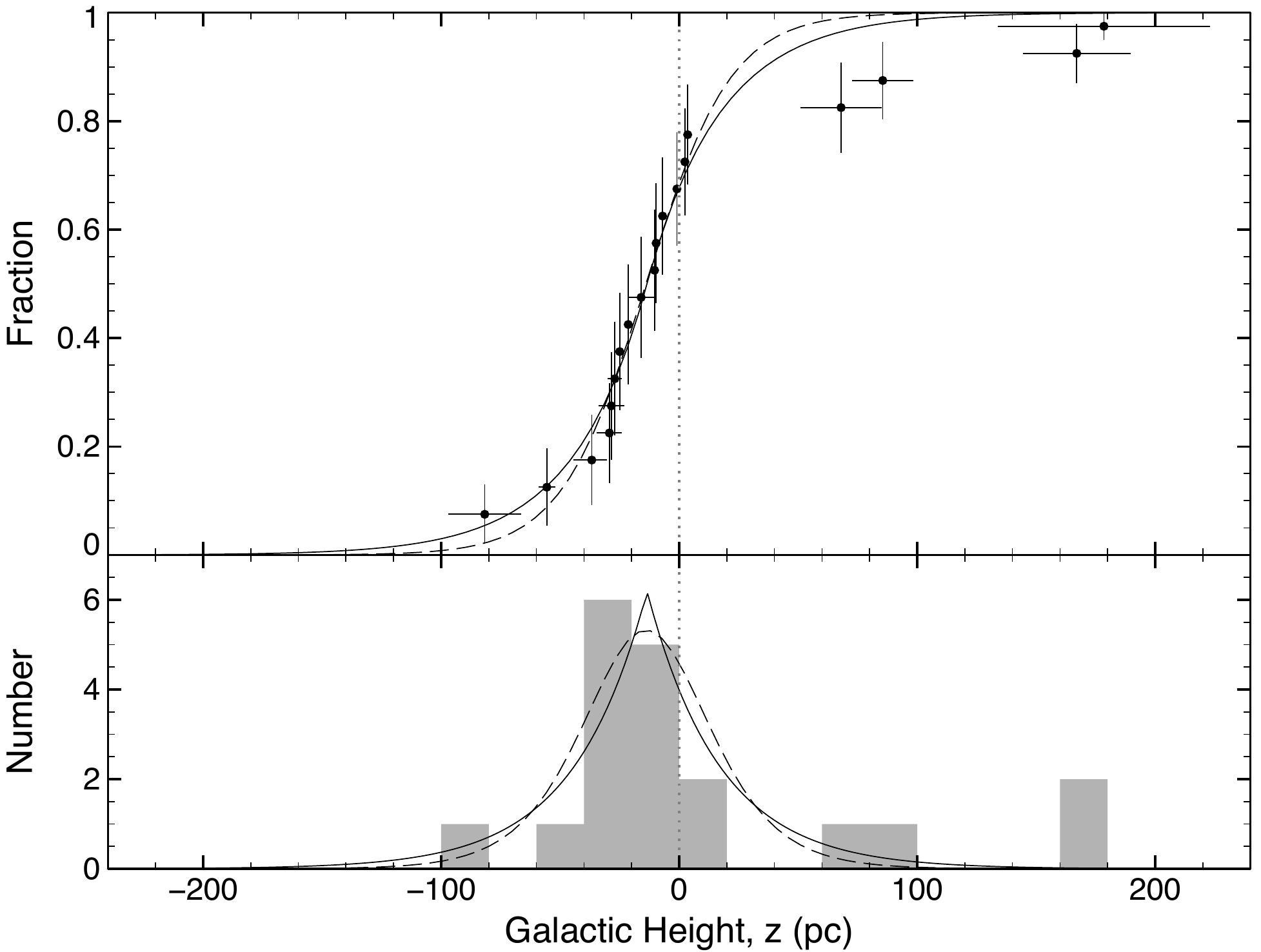}
\end{minipage}
\caption{Left:  Top-down view of the Galaxy, with the Galactic Center at  (0,0) and the Sun marked by a cyan arrow. 
The grayscale shows the distribution of free electrons \citep{cl01}. Known magnetars are shown as red circles with 
distance uncertainties indicated, known X-ray Isolated Neutron Stars (XINSs) are shown by yellow circles, and 
all other known pulsars are blue dots.  Right:  Top panel: Cumulative distribution function of the height z above the Galactic plane for the 
19 known magnetars in the Milky Way. Data are fit to an exponential model (solid line) and a self-gravitating, isothermal disc 
model (dashed line).  Bottom panel: Histogram of the distribution in z of the known Galactic magnetars. Lines are as above.
Note the offset of the peak from zero; this is the offset of the Sun from the Galactic Plane.  From OK14.}
\label{fig:galdist}
\end{figure}

The inferred spatial locations within the Galaxy (see Fig.~\ref{fig:galdist}) strongly suggest we are biased
against finding more distant magnetars. 
%presumably because bursts (the phenomenon most likely
%to result in detection given existing technology) from more distant objects go undetected.  
Nevertheless, some inferred
distances that are comparable to that of the Galactic Center indicate that
we have been sensitive to a large fraction of the volume of the Milky Way; future all-sky monitors
with only modest increases in sensitivity could fully flesh-out the active magnetar population in the Galaxy.

\subsection{Associations with Supernova Remnants and Wind Nebulae}
\label{sec:snrs}

Of the 23 confirmed magnetars, 8 are reliably associated with supernova remnant shells and an additional
2 have possible associations.  The large number of remnant associations is fully consistent with the great youth implied
both by magnetar spin-down ages, as well as by their proximity to the Galactic Plane.  
The associated remnants lack unusual properties when compared
with shell remnants that harbor neutron stars with lower magnetic fields \citep{vk06,mrtp14}.   
This appears to be in conflict with the proposal of \citet{dt92a} that
magnetars form from neutron stars rotating with period $\sim$1~ms at birth, which assists a fast dynamo.
%Indeed detailed simulations have shown that a millisecond magnetar could power observed hyper-luminous supernovae and gamma-ray bursts
%\citep{mgt+11,bmtq12}.
The difficulty with this picture is that a neutron star with magnetic field $> 10^{14}$~G spinning at 1 ms 
quickly loses most of its rotational energy, releasing energy
in excess of $10^{52}$ erg, greater than the supernova explosion energy itself.  It is therefore likely to be associated with
either anomalously large shell remnants, or else no remnant at all, if it expanded sufficiently rapidly to dissipate on a few hundred year time scale.
%As magnetar remnants show no difference from regular remnants, the dynamo origin of magnetar fields put forth by \citet{dt92a} has been 
%called into question, and 
The normality of magnetar supernova remnants challenged the dynamo model and led to discussion of strong fossil fields from the progenitor star \citep{fw06},
however the latter is not without difficulties \citep[see, e.g.][]{spr08}.

%\subsubsection{Evidence for `Magnetar Wind Nebulae'}
%\label{sec:mwn}

Extended, nebular emission near magnetars, `magnetar
wind nebulae (MWN),' may exist, in analogy with pulsar wind nebulae (PWNe).
PWNe are extended synchrotron
nebulae surrounding some radio pulsars, a result of the interaction of relativistic,
magnetized pulsar particle winds interacting with their environments, and
ultimately powered by the pulsar's rotation \citep[see][for
a review]{gs06}.  Observations
of a MWN could, in principle, provide important information on the
composition and energetics of continuous particle outflows from magnetars.
Clear evidence for temporary magnetar outflows has been seen in the form of transient extended radio emission following two giant flares \citep{fkb99,gkg+05,gle+05}.
%\citet{fkb99} observed a fading, albeit unresolved, radio source following the 1998 giant flare from SGR 1900+14 which the authors argued was due
%to a synchrotron nebula produced by particle outflow in the giant flare.  \citep{gkg+05} and \citep{gle+05}
%reported on a relativistically expanding linearly polarized radio nebula observed using the Very Large Array and the Australia Telescope Compact
%Array after the 2004 giant flare from SGR 1806$-$20.  Those authors
%argued the nebula resulted from shocked ambient medium following the release of $\sim 10^{24.5}$~g of material having initial kinetic
%energy $\sim 10^{44.5}$~erg.  \citet{tgg+05} detected expansion directly in Very Large Array data, inferring an expansion velocity of 0.26c followed by deceleration
%simultaneous with a radio re-brightening, which together suggest the onset of a Sedov-Taylor phase.
However interesting, these do not constitute MWNe since the latter by definition are long-lived and result
from continuous particle outflow even when the magnetar is in quiescence.

There are multiple reports of stable MWNe in the literature \citep[e.g.][]{rmg+09,crb+13},
%The first possibility was surrounding not a magnetar but a Rotating Radio Transient (RRAT), J1819$-$1458.   This object is a sporadic
%radio emitter whose potential relevance to magnetars is its long spin period (4.3 s), its relative youth (characteristic age 117 kyr), and
%its high spin-inferred magnetic field, $5 \times 10^{13}$~G \citep{mll+06}.
%\citet{rmg+09} and \citet{crb+13} reported possibly related extended X-ray emission around J1819$-$1458, but noted a large ratio of X-ray to spin-down
%luminosity for the nebula ($\eta \equiv Lx/\dot{E} \simeq 0.2$), much higher than seen for PWNe, and suggested it may be powered somehow
%by magnetic energy.  They could not, however, rule out the extended emission being a dust-scattering halo.
%Indeed, \citet{vb09} reported an X-ray MWN around the magnetar 1E~1547.0$-$5408, but \citet{okn+11} showed the flux of the extended emission
%was strongly correlated with the magnetars' flux, concluding it must be a dust scattering halo.
however extended emission can also be due to dust scattering along the line of sight \citep[e.g.][]{okn+11}.
Currently the most compelling case of a MWN is an asymmetrical X-ray structure around Swift J1834.9$-$0846 \citep{ykk+16}.
One thing is certain:  
%even if one or more of the above putative MWNe were real, 
the phenomenon is not generic to the source class.  
Deep X-ray imaging observations of many different
magnetars have revealed no evidence for any MWN down to constraining (though, unfortunately, not generally well quantified) 
limits \citep[e.g][]{ghbb04,akac13}.  
%Why some magnetar have nebulae and others do not is unclear and will require more MWN detections to be answered.
However, \citet{hg10} report a potential association between a magnetar, CXOU J171405.7$-$381031, in 
the supernova remnant CTB 37B, and 
TeV emission, which they speculate may be a relic MWN.

%Mention Wind braking  Tong et al. 2013?

%\citet{ykk+12} reported on two {\it XMM-Newton} observations of magnetar Swift J1834.9$-$0846, one in 2005, well prior to this source's 
%2011 X-ray outburst, and one month afterward.
%They find both a symmetric X-ray ring surrounding the point source, argued to be a dust-scattering halo, as well as asymmetrical X-ray emission 
%present at approximately the same level in both observations.
%The efficiency $\eta \simeq 0.7$ for the asymmetrical emission's X-ray luminosity, again higher than for PWNe, suggesting an alternative energy 
%source likely related to the high magnetic field inferred
%for the central source from its spin-down, $1.4 \times 10^{14}$~G.  \citet{etr+13} suggested the asymmetrical feature
%is due to dust scattering by the same cloud
%producing the symmetrical structure, because of some variability seen between epochs.  However very recently, \citet{ykk+16} 
%reported on two more recent {\it XMM} observations long post-outburst, showing the putative wind nebula remains essentially
%unchanged.  This is strong support for the reality of the nebula.

\subsection{Relationship to High-Magnetic-Field Radio Pulsars}
\label{sec:highB}

%{\bf LEAVE THIS SECTION FOR CLOSER TO THE END OF THE PAPER?}
If high magnetic fields in neutron stars are responsible for the dramatic X-ray and soft-gamma-ray activity in magnetars, and given
that some magnetar behavior has been seen in apparently low-B sources \citep[such as SGR 0418+5729;][]{ret+10}, then it stands
to reason that high-B radio pulsars may occasionally exhibit magnetar-like activity \citep{km05,nk11}.  
This possibility was also hinted at by higher blackbody temperatures in high-B radio pulsars compared
with lower-B sources of the same age \citep[e.g.][]{zkm+11,ozv+13}.
The idea of high-B radio pulsars as quiescent magnetars has proven
to be correct.

The first example came from the 
young (age $<$1 kyr), high-B ($5 \times 10^{13}$~G) but curiously
radio undetected \citep{aklm08} rotation-powered pulsar 
PSR J1846$-$0258 in the supernova remnant Kes 75 \citep{gvbt00}.   In long-term monitoring designed to 
measure the source's braking index, \citet{ggg+08} detected a sudden X-ray outburst lasting 
$\sim$6~wks and
of total energy $\sim 3 \times 10^{41}$~erg in the 2--10-keV band \citep[see also][]{ks08b}.  The outburst also included
several short magnetar-like bursts and a large glitch with unusual recovery \citep{kh09,lkg10}.
%This was the first example of a temporary metamorphosis of an otherwise \citep[though curiously radio-quiet][]{aklm08}
%ordinary pulsar into a magnetar-like object.   
Post-outburst, the source has returned to its quiescent, apparently rotation-powered
state, albeit with enhanced timing noise and a significant change in braking index, from 
2.65$\pm$0.01 pre-outburst to 2.19$\pm$0.03 post-outburst \citep{lnk+11,akb+15}.  
%So large a change in braking index
%is unprecedented for a rotation-powered pulsar and poses a challenge to models of pulsar spin-down.
A second such metamorphosis was seen very recently in PSR J1119$-$6127, a {\it bona fide} radio pulsar, also
very young (age $<$ 2 kyr) and apparently high-B ($4 \times 10^{13}$~G).  In this case the outburst was
heralded by bright, magnetar-like X-ray bursts \citep{ykr16,klm+16,glk+16}.  Follow-up X-ray observations \citep{akts16}
showed an increase in X-ray flux of nearly a factor of 200 with a dramatic spectral hardening.
%, similar to what is observed in magnetar outbursts (see \S\ref{sec:outbursts}).  
This outburst was also accompanied
by a large glitch \citep{akts16} and, remarkably, by a temporary cessation of radio emission \citep{bpk+16}.
%It seems likely a change in braking index will be seen post-outburst.  
%Interestingly, prior to this tell-tale event,
%PSR J1119$-$6127 had shown unusual radio variability and glitch recoveries \citep{wje11,awe+15} which
%were distinctly magnetar-like.

The similarity of the PSR J1846$-$0258 and PSR J1119$-$6127 events both to each other and to magnetar
outbursts, together with the fact that they are both relatively rare high-B rotation-powered pulsars normally,
confirms the close relationship between radio pulsars and magnetars, and
the correlation between high spin-inferred magnetic field and magnetar-like activity.   
%That magnetar-type emission
%comes nearly exclusively from high-spin-inferred magnetic field sources is clearly no coincidence.

The discovery of radio pulsations from magnetars \citep[see \S\ref{sec:multitemporal}][]{crh+06,crhr07} also 
provides an observational link between radio
pulsars and magnetars.  However, the radio properties of magnetars
(see \S\ref{sec:multitemporal} and \S\ref{sec:spectra_oir}) are somewhat different from
those of conventional radio pulsars.

\section{TEMPORAL BEHAVIOR}
\label{sec:temporal}

%\section{Temporal Behavior}

\subsection{X-ray Pulsations, Spin-Down and Timing}
\label{sec:pulsations}
% MONDAY

%A basic observational property of magnetars is their X-ray pulsations.
%In the known magnetar population, pulsation periods lie roughly uniformly in the range 2--12 s. 
%The narrowness of this range itself is interesting and is discussed below.  
%Note however that
%magnetar-like behavior has been seen from two apparently rotation-powered pulsars
%having much shorter spin periods, 0.327 s \citep{ggg+08} and 0.407 s \citep{akts16}.
%Also, very recently, strong evidence for a 6.67-hr rotation period in a magnetar has been
%presented.  This is discussed further below.

\begin{figure}
\begin{minipage}{2.9in}
\hspace{-0.2in}
\includegraphics[scale=0.3]{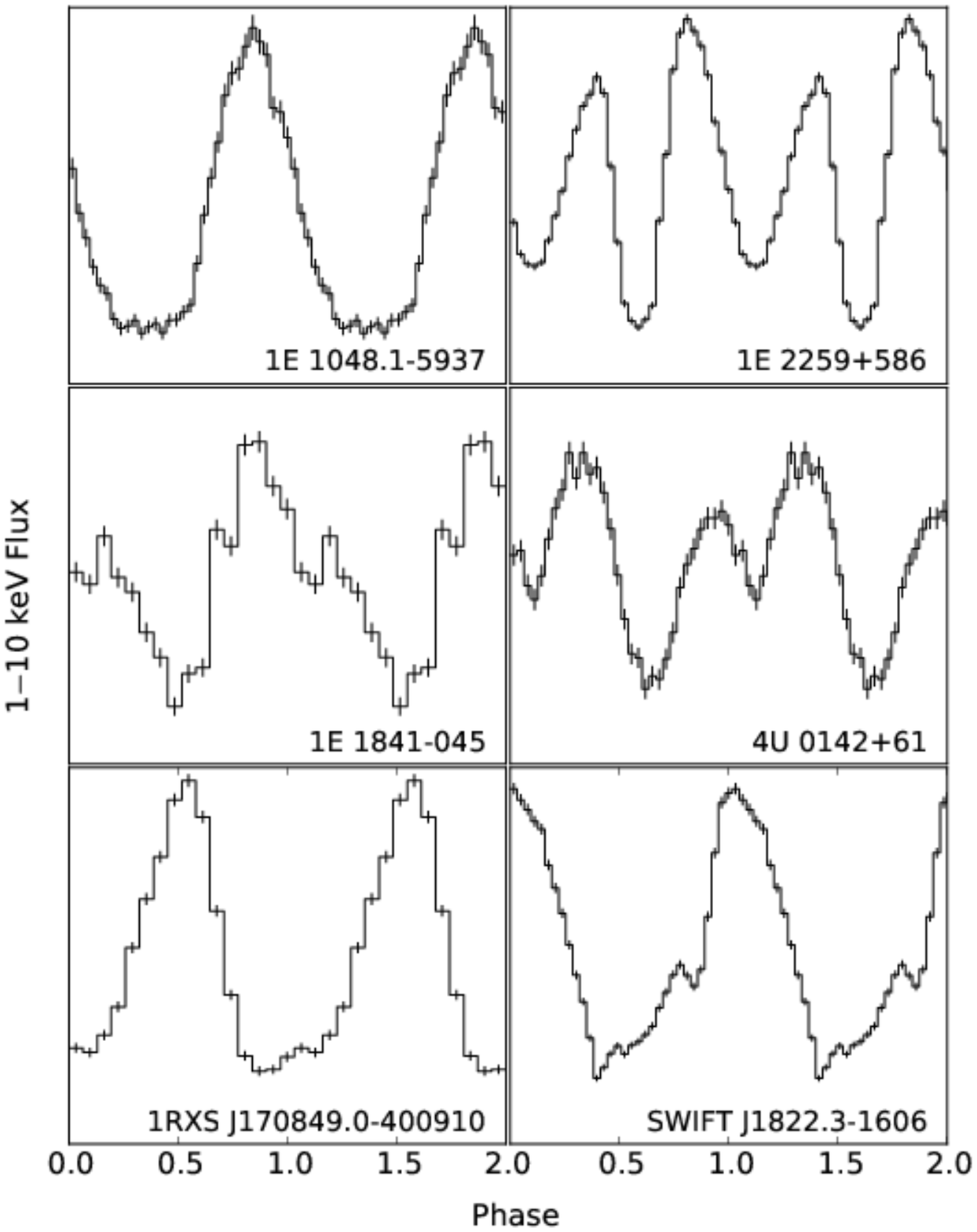}
\end{minipage}
\hfill
\hspace{-0.55in}
\begin{minipage}{2.9in}
\vspace{0.4in}
\includegraphics[scale=0.13]{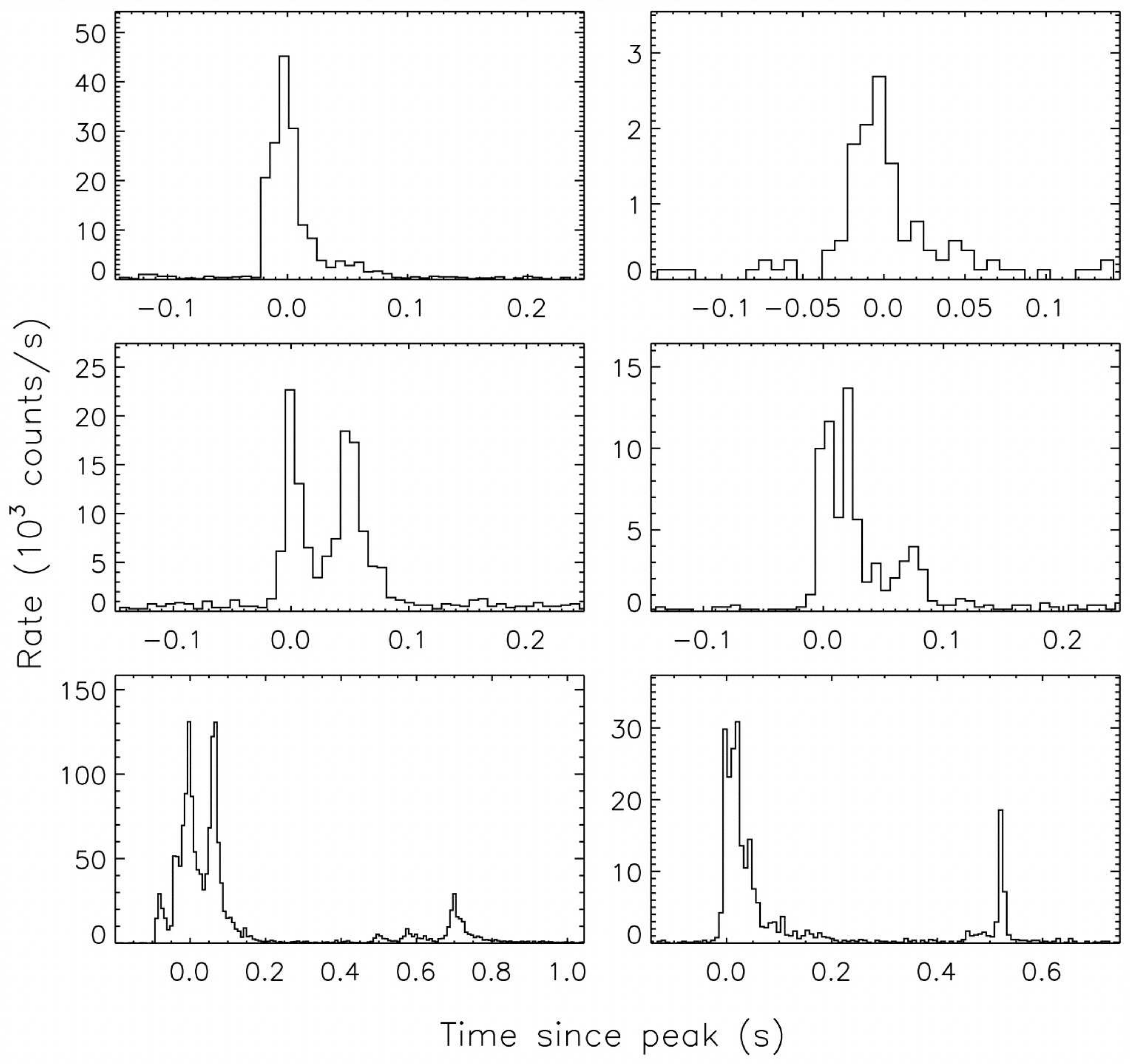}
\end{minipage}
\caption{(Left) Several X-ray pulse profiles of magnetars in the 1---10-keV band (courtesy R. F. Archibald).
(Right) Examples of single bursts from SGRs 1806$-$20 and 1900+14 shown with 7-ms time resolution in
the 2--60-keV band from {\it RXTE}/PCA data.  From \citet{gkw+01}.
}
\label{fig:xrayprofiles}
\end{figure}

Magnetar X-ray pulse profiles are generally very broad, with one or two components, and
with duty cycles that approach 100\% (see Fig.~\ref{fig:xrayprofiles} left).  X-ray 
pulsed fractions (the fraction of point-source emission that is pulsed)
are typically $\sim$30\% but range from $\sim$10\% \citep{ioa+99} to as high as $\sim$80\% \citep{kkp+12}.
The profile morphologies and pulsed fractions can vary strongly with energy.
Pulse profiles are generally stable apart from during outbursts
where large profile variations are common.
However there is at least one example of long-term, low-level pulse
profile evolution, in magnetar 4U~0142+61 \citep{gdk+10}.

Because the X-ray pulsations are generally the most practically observable,
long-term timing of magnetars has been done almost exclusively therein, with
some inclusions of radio data where available (see \S\ref{sec:multitemporal}).  
In the past, when X-ray telescope time was very difficult to acquire due to required
long exposures, timing was
done by measuring the pulse period at multiple epochs, typically spaced by months
to years \citep[e.g.][]{bs96,bss+00}.  This clearly demonstrated regular spin-down in
the first known sources, which distinguished them from the accreting
X-ray pulsars as these often spin up.

Today, the timing method used most commonly
is ``phase-coherent timing,'' borrowed from the radio pulsar world 
and brought first to the magnetar world thanks to the great sensitivity and ease of scheduling
of {\it RXTE} \citep{kcs99}.  In this technique, 
{\it every} rotation of the pulsar is accounted, sometimes over years, enabling precise measurements
of period and eventually spin-down rate.  Phase-coherent timing has been accomplished
long-term for the 5 brightest known magnetars using {\it RXTE} \citep[see][and references therein]{dk14}
and now {\it Swift} \citep[e.g.][]{akn+13}.  Phase-coherent timing has also sometimes been
done following magnetar outbursts, however pulse profile changes,
common in outbursts, make this difficult.  Moreover, as the source fades, particularly for
the `transient' magnetars that are faint in quiescence,
ever-longer integration times are required, often rendering long-term timing impractical.

%\begin{textbox}
%In phase-coherent timing, carefully spaced snapshots of the source are obtained, with integration
%times just long enough to determine the phase $\phi$ of the pulse, typically to within at least $\phi \simeq 0.1$.
%A stable pulse profile is crucial to permit a reliable fiducial point -- presumably
%corresponding to a fixed location on the neutron star -- for phase measurement.
%Given even a rough initial period (say from a periodogram in a single observation), 
%the initial spacing of the first two snapshots, typically separated by several hours,
%determines an improved period, thanks to the long time baseline.  With the next observed typically
%the next day, the improvement on the pulse period is improved further, thanks to the extended
%baseline.  With each improvement, the time baseline until the next snapshot can be extended.
%This works well until a period derivative begins to matter, which introduces a quadratic term
%into the phases, since
%\begin{equation}
%\phi(t) = \phi(t_0) + \nu (t-t_0) + \frac{1}{2}\dot{\nu} (t-t_0)^2,
%\end{equation}
%where the frequency $\nu \equiv 1/P$ with $P$ the pulse period.
%This introduction of the quadratic enables an initial meaurement of the spin-down rate $\dot{nu}$
%typically after only a few days of observations for magnetar-strength fields.  
%Once determined in this way, barring a glitch or large rotational instabilities,
%the phase can be monitored and modelled with only monthly snapshots, leading to
%precise values for $\nu$ and $\dot{\nu}$.
%\end{textbox}

Phase-coherent timing has enabled very precise measurements of $P$ and $\dot{P}$ 
now for many magnetars; by the
same token, however, deviations from simple spin-down become easily detectable with this technique.
In this regard, magnetars are quite prolific.  Ubiquitous in their rotational evolution is what is termed
``timing noise:'' apparently random wandering of phases particularly
on months to years time scales.  Such wandering is phenomenologically modelled by higher-order derivatives
of $P$ (or, equivalently, $\nu$) and in some cases, as many as 10 such derivatives are required \citep[e.g.][]{dk14}.
%Nevertheless, even given all the observed spin-down noise, in no case has any extended episode of spin-up been observed.
However, apart from the smoothly varying wander of timing noise,
and thanks to the available timing precision inherent to phase-coherent timing, 
sudden spin-ups (``glitches'') (and, more rarely, anti-glitches) are now observed frequently in magnetars
(\S\ref{sec:glitches}).

%Mention low-Pdot magnetars here?
% no in overall section

%One interesting question is regarding the origin of the observed narrow magnetar period range.  
Whereas most radio pulsars are thought to be born with periods
of at most a few hundred milliseconds, that the shortest known {\it bona fide} magnetar
has a relatively long 2-s period in spite of a young age 
is surely a result of rapid magnetic braking. 
%inevitable in so highly magnetized a neutron star.
%Faster rotators may exist, but they slow down so rapidly (in under $\sim$1000 years), they are rare to observe.
The long period cutoff of 12~s (but see below) has been more of a puzzle, which is related to the life-time
of magnetar activity \citep[e.g.][]{cgp00,vrp+13}.  

Finally, one source shows unusual timing behavior that is worthy of note.
1E~1048.1$-$5937 has shown episodes of large (factor of 5--10) spin-down rate variations which appear to be,
curiously, quasi-periodic on a time scale of $\sim$1800 days \citep{akn+15}.  These episodes have, in all cases
observed thus far, followed major flux outbursts, with a delay between the radiative outburst and spin-down fluctuations
of $\sim$100 days.  
%The origin of the quasi-periodicity is not known, but with only 3 such events observed, one may question
%whether indeed it is real.  
A fourth such flux outburst has very recently begun, at the approximate epoch
predicted by the apparent quasi-periodicity \citep{atsk16}. 
%The interested reader is encouraged to watch the literature
%to see if the trend indeed persists.  A fourth such event would be difficult to dismiss and would require some physical explanation.
%One possible origin is suggested by \citep{akn+15} to lie in a magnetospheric instability;  the rationale
%for this is the apparent similarity to other curious long-time scale variations in the spin-down rates and radio pulse profiles
%of some radio pulsars \citep{lhk+10}.

\subsubsection{Glitches}
\label{sec:glitches}

Phase-coherent timing of magnetars by {\it RXTE} enabled the discovery that magnetars are among
the most frequently glitching neutron stars known \citep{klc00,dkg08}.  A ``glitch,''  a phenomenon common to
young radio pulsars \citep[e.g.][]{ymh+13}, consists of a sudden spin-up, typically involving $\Delta\nu/\nu$ in the range $10^{-9} - 10^{-5}$
in both magnetars and radio pulsars.  
%Note however for the same fractional change, since magnetars rotate so much slower, the effective
%change in angular momentum during a magnetar glitch is $\sim 100\times$ smaller.
Also typically associated with glitches in both radio pulsars and magnetars are long-term changes in spin-down rate $\dot{\nu}$, with
typical $\Delta\dot{\nu}/\dot{\nu}$, almost always positive, of at most a few percent and usually far smaller in radio pulsars.
A common phenomenon is glitch recovery, in which a sizable fraction of the glitch (in radio pulsars, 
typically 0--0.5) recovers quasi-exponentially within a week or two following the glitch.
A remarkable behavior seen practically exclusively in magnetars is extremely strong glitch recovery, such that the full
initial spin-up is recovered, and in some cases {\it over}-recovery is seen, such that the overall effect is a spin-{\it down}
\citep[e.g.][]{gdk11}.
These strong recoveries involve initially very large values of $\dot{\nu}$, sometimes upwards of 10$\times$ the 
pre-glitch long-term $\dot{\nu}$ \citep[e.g.][]{kgw+03,kg03,dis+03,wkt+04}. 
Additionally, in one magnetar-like glitch case, the level of timing noise was observed to be greatly
enhanced for several years following the over-recovered glitch \citep{lnk+11}.

%What is the origin of the remarkable glitch recoveries seen in magnetars?  The answer to this question is presently unknown.
%However, if, during the glitch recovery, the external torque on the neutron star remains constant, then
%a very large fraction of the moment of inertia of the star (which of necessity involves the stellar core)
%must decouple from the crust at the glitch epoch, then slowly re-couples.  
%This is not formally ruled out, however 
The coincidence of many such glitches
and recoveries with large radiative outbursts and their relaxations is suggestive of magnetospheric phenomenon. 
Indeed many -- though not all -- magnetar glitches occur at epochs of large X-ray flux outbursts, which themselves
are often accompanied by many outward radiative changes such as short bursts and X-ray pulse profile changes (see \S\ref{sec:outbursts} below).
Curiously some magnetars have only shown radiatively silent glitches \citep[e.g. 1RXS J170849.0$-$400910; ][]{sak+14} and some have shown both
silent and loud glitches \citep[e.g. 1E~2259+586; ][]{dk14}.  
%It could be that glitches are an internal phenomenon that can trigger magnetospheric
%twists, depending on, for example, depth.  This is discussed further in \S\ref{sec:andrei}.

Several apparent ``anti-glitches'' have also been reported in magnetars
\citep{wkv+99,sag14}.
These events appear consistent to within the available time resolution with being sudden spin-{\it downs}  of magnetars
and have not been seen in radio pulsars at all.  The most convincing of these 
is the anti-glitch reported in 
1E~2259+586 \citep{akn+13} in which an apparent sudden spin-down of amplitude $\Delta\nu/\nu \sim 10^{-7}$ accompanied
a bright, short X-ray burst and a long-lived flux outburst.  Although this event could in principle have resulted from
an over-recovered spin-up glitch, the recovery time scale of the initial event would have had to be at most a few days, much shorter than
any previously observed glitch recovery.  
%Actually, the 1E~2259+586 anti-glitch was not the first reported in a magnetar.
%Previously a possible very large ($\Delta\nu/\nu = 1 \times 10^{-4}$) anti-glitch was reported in magnetar
%SGR 1900+14 \citep{wkv+99} near its 1998 giant flare.  However the spin-down occured in an 80-day observing gap, 
%so could not be demonstrated to have been a truly sudden event.  Also,
%\citet{sag14} reported an anti-glitch in magnetar 1E 1841$-$045 however there was no evidence for any radiative
%enhancement at the relevant epoch and an analysis of the same data by \citet{dk14}
%does not find an such anti-glitch.  
The origin of anti-glitches is still debated; see \S\ref{sec:andrei} for further discussion.

\subsection{Transient Radiative Behavior: Bursts, Outbursts, Giant Flares \& QPOs}
\label{sec:transient}

%The original hallmark observational characteristic of a magnetar is
%its dramatic X-ray and soft-gamma-ray bursting and flaring.  Magnetars,
%when active, produce sudden radiative outbursts that span orders of
%magnitude in temporal and luminosity space.  
%Some nomenclature introduction is in order.  
The term ``bursts'' here is used to mean the short, few millisecond to
second events, some of which are followed by longer-lived ``tails,'' an afterglow
of sorts.  The term ``outburst'' is used to describe a sudden but much longer-lived
(weeks to months) flux enhancement, which typically involves many of the shorter
bursts, and involves a long (many months) ``tail'' or afterglow.
The term ``giant flare'' is reserved exclusively for what appear to be catastrophic
events involving the sudden release of over $10^{44}$~erg of energy.
%These events are the ones that saturate detectors and invoke superlatives like
%`greatly outshone the entire cosmic hard X-ray sky for a few moments' or
%`released as much energy in a few seconds as does the Sun in 250,000 years.'
``Quasi-periodic oscillations'' have been seen in the tails of some giant flares.
It is fair to say that nearly all magnetar radiative variability unrelated
to their pulsations can be placed into one of these categories.

\subsubsection{Bursts}
\label{sec:bursts}

Short bursts are by far the most common type of magnetar radiative event.
There are magnetars which have emitted thousands
of bursts, usually very much clustered in time, and there are magnetars that, in spite
of intensive monitoring programs, have shown at most a handful of bursts.
Indeed the former were long thought to be the `SGRs' of the magnetar population,
with the latter being the `AXPs.'  However with further study it appears that 
there is a full spectrum in burst rates, and some sources that might originally
have been thought to be extremely active (e.g. SGR 0526$-$66) have lain dormant
for decades subsequently \citep{kkm+03,tem+09}.  Similarly, sources that for years were not known to burst
(e.g. 1E~2259+586) suddenly entered an active burst phase, emitting several dozen
bursts in a few days \citep{kgw+03}.  This is one key reason the AXP/SGR classification scheme
seems obsolete.  Note that bursts are more common during outbursts (see \S\ref{sec:outbursts})
however there are examples of bursts occuring when the source appears otherwise in
quiescence \citep[e.g.][]{gkw02}.

There have been multiple detailed statistical studies of the properties of
short magnetar bursts \citep{gwk+99,gwk+00,vkg+12,lgkk13}.
Here we summarize the findings of these studies.
Burst peak luminosities can be hyper-Eddington but span a broad spectrum, typically
ranging from $10^{36}$ to $10^{43}$~erg~s$^{-1}$, with bursts detected
right down to the sensitivity limit of current X-ray detectors.  
Burst durations span over two orders of magnitude, ranging between a few ms and a few sec,
with distributions typically peaking near $\sim$100 ms.
Burst fluence distributions are generally well described with power-law functions
of indexes in the range  $-$1.6 to $-$1.8 \citep{gwk+99,gwk+00}.
Bursts are usually but not always
single-peaked, with the rise typically faster than the decay.
Interestingly, although some studies have shown that short bursts arrive
randomly in pulse phase \citep[e.g.][]{gwk+99,gwk+00}, others have found a preference
for bursts near the pulse maximum \citep[e.g.][]{gkw04}.
Figure~\ref{fig:xrayprofiles} (right) shows examples of short magnetar bursts.

Some bursts show long, sometimes several-minute tails \citep[e.g.][]{lwg+03,gwk+11,mgk+15}, during which the pulsed
flux is sometimes greatly enhanced \citep[e.g.][]{wkg+05,mgw+09,akb+14}.  In this way they are almost like miniature giant flares
(see \S\ref{sec:giantflares}).  Tails fade slowly, with decays well described by relatively flat power laws
of indexes well under unity.  Though much lower in flux, the long-duration tails
can sometimes contain significantly more energy than the burst itself.  Ratios of
burst to tail energies can vary by over an order of magnitude in different sources and even in
different bursts in the same source \citep[e.g.][]{gwk+11}.

%Observationally, it has been suggested \citep{wkg+05} that there are two classes
%of bursts: those with long tails in which tail energy greatly exceeds burst energy, in which 
%This dichotomy has been suggested to align with the predicted picture of
%\citet{lyu0?} 

\subsubsection{Outbursts}
\label{sec:outbursts}

\begin{figure}
\begin{minipage}{2.9in}
\includegraphics[scale=0.28]{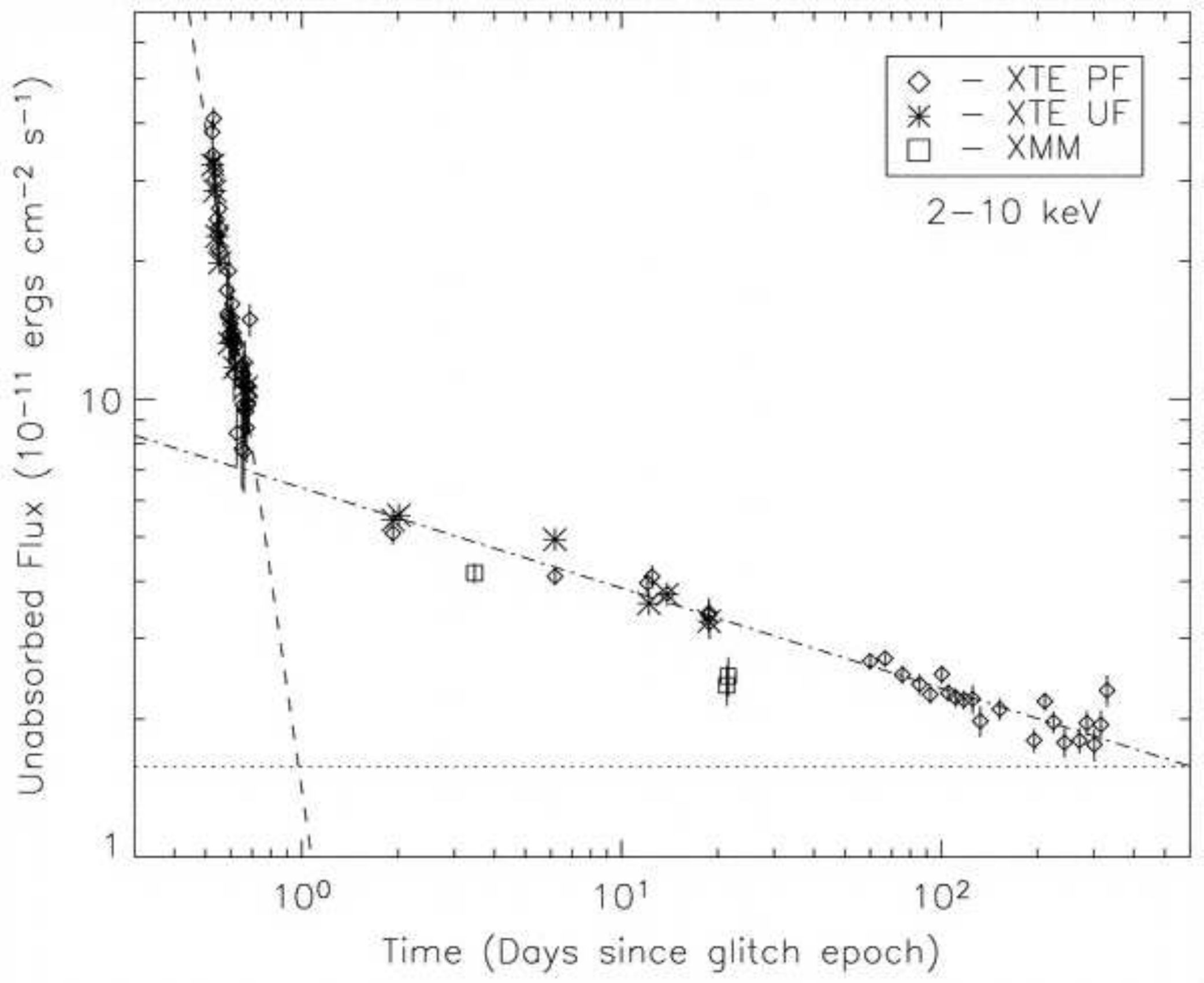}
\end{minipage}
\hfill
\hspace{-0.75in}
\begin{minipage}{2.9in}
\hspace{.35in}
\includegraphics[scale=0.12]{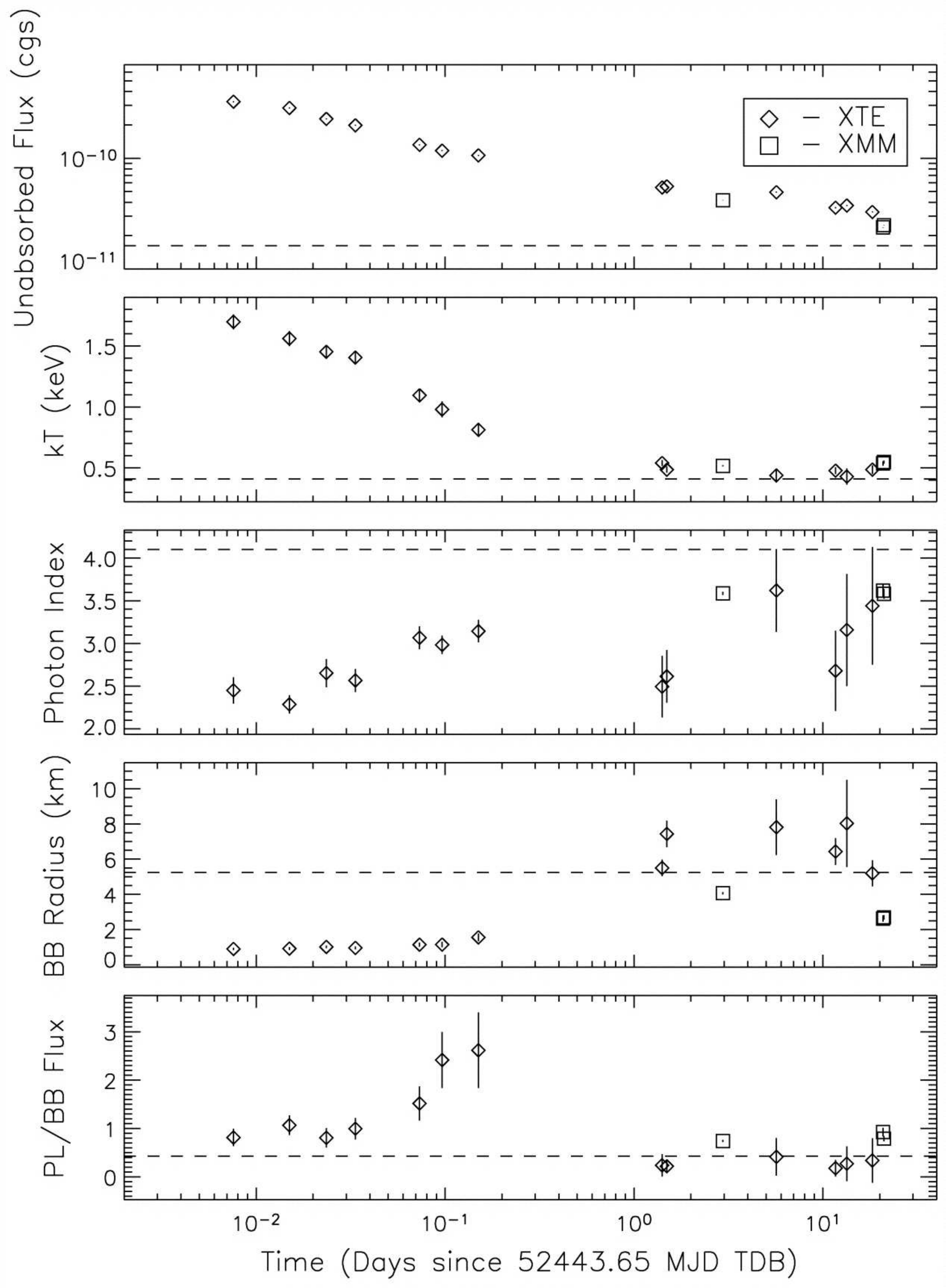}
\end{minipage}
\caption{(Left) Flux evolution during and after the outburst of 1E 2259+586.
Note the initial rapid decrease in flux on the first day, when the vast majority of associated short bursts were detected, followed by the slower
subsequent fading. 
(Right) Spectral evolution of 1E 2259+586 through and following its 2002 outburst (discussed in \S\ref{sec:specevol}).
Top to bottom: unabsorbed flux (2–10 keV), blackbody temperature (kT),
photon index, blackbody radius, and ratio of power-law (2–10 keV) to bolometric blackbody flux.
Horizontal dashed lines denote the values of each parameter fortuitously measured
one week prior to the outburst.  From \citet{wkt+04}.
}
%(Right) Compilation of post-outburst light curves of magnetars showing
%the variety of flux relaxation behaviors.
%From \citet{re11}.}
\label{fig:outburst}
\end{figure}

A magnetar `outburst' is an event consisting of a large (factor of 10--1000) and usually sudden increase in
the source X-ray flux, sometimes as high as $10^{36}$~erg~s$^{-1}$ \citep[see][for a compilation]{re11}.  These events are generally 
accompanied by a bevy of other radiative anomalies such as spectral hardening, change in pulsed fraction, pulse profile changes 
(often from a simpler to a more complex profile), multi-wavelength changes, and multiple short X-ray bursts.
Most outbursts for which there are available data also are accompanied by some form of timing
anomaly, ususally a spin-up glitch or occasionally an anti-glitch.  The flux following an outburst
usually decays on multiple time scales, with a very rapid initial decay within minutes to hours
(and hence which is often missed by observatories) followed by a slower decay \citep[e.g.][]{wkt+04},
sometimes termed an `afterglow' which can last months to years.  These slowly fading afterglows are often quasi-exponential
\citep[e.g.][]{rit+09} but sometimes have an interesting
time evolution, with periods of power-law decays interrupted by few-month periods of flux stability \citep[e.g.][]{akt+12}.
In general, magnetar outbursts (even those from the same source) show a variety of time scales for their 
relaxations \citep[e.g.][]{etr+13}.
As in the tails of the more common bursts, there is a diversity in the ratios of burst to afterglow energies, ranging from a
few percent to one to two orders of magnitude \citep{wkt+04}.  
An example of an X-ray light curve from a magnetar in outburst
is shown in Figure~\ref{fig:outburst}.
A description of the spectral evolution of magnetars during and following outbursts is
deferred to \S\ref{sec:spectral} below.
%The aforementioned strange quasi-periodic outbursts of 1E~1048.1$-$5937 are also notable for their slow rises;
%most magnetar outbursts begin suddenly.

Some sources have shown no outbursts over decades \citep[e.g. 1RXS J1708$-$4009; ][]{dk14}
while others
have had multiple \citep[e.g. SGRs 1806$-$20, 1900+14, 1E 1547.0$-$5408;][]{wkf+07,ggo+11,nkd+11}. 
In the past, AXPs have tended to be associated with sources
that have few if any outbursts, whereas SGRs are sources that are much more outburst-active.  However in terms of
reasonable outburst rate estimates, there is no clear evidence for bi-modality, suggesting a full continuum of
activity.  
%Indeed a continuum of behaviors is predicted by current models of magnetothermal evolution and
%depends on source age and magnetic field \citep[][see \S\ref{sec:andrei}]{pp11a}.

%Note that the known magnetar population is likely heavily biased toward
%sources that are more outburst-active, since this is very frequently how they are discovered.  
%This is thanks to sensitive all-sky monitors built to study gamma-ray bursts, such as
%the Burst Alert Telescope aboard {\it Swift} and the Gamma-Ray Burst Monitor aboard {\it Fermi}.
%Indeed \citet{ok14} showed a clear increase in the rates of magnetar discoveries at the times
%of the launches of these two telescopes.   
The term `transient' magnetar was introduced to describe those sources which have very low
$< 10^{33}$~erg~s$^{-1}$ quiescent luminosities, which generally go unnoticed until they produce
outbursts involving flux increases of factors of 100-1000, accompanied by bright bursts that 
trigger monitors \citep[e.g.][]{mgz+13}.
The first discovered transient magnetar was XTE J1810$-$197.  It was caught in an outburst in 2003
\citep{ims+04} and observed to decay on a year time scale
\citep{gh07}. 

Transient magnetars may in fact be the norm among the magnetar population, with the
well studied bright sources like 4U~0142+61 or 1E~2259+586 unusual for relatively high
quiescent luminosities, upwards of $10^{35}$~erg~s$^{-1}$.  Why some magnetars in quiescence
are much brighter than most is an interesting question.
Some transient magnetars, e.g. SGR 0418+5729 \citep{ret+10} or Swift J1822.3$-$1606
\citep{rie+12,skc14}, have low spin-inferred dipole magnetic fields and are thought to possess
strong internal toroidal fields.

%This also is a plausible explanation for the apparently low-magnetic-field magnetars like
%SGR 0418+5729 \citep{ret+10} or Swift J1822.3$-$1606 \citep{rie+12,skc14}. 
%(see \S\ref{sec:pulsations} above).
%This is discussed in more depth in \S\ref{sec:andrei}.

%Further,
%some magnetars ever never reach a truly `quiescent' flux state given that sometimes
%new outbursts occur before a full relaxation from the previous one \citep[e.g.][]{nkd+11}.  On the other hand, at least the five
%brights sources monitored over 16 years with {\it RXTE}, and now with {\it Swift}, have generally stable
%fluxes for years after any outburst.  

\subsubsection{Giant Flares}
\label{sec:giantflares}

The queen of magnetar radiative outbursts is the giant flare.
Thus far, only three giant flares (GFs) have been recorded, all from
different sources.  These events occured on March 5, 1979 \citep[SGR
0526$-$66; ][]{ekl+80}, August 27, 1998 \citep[SGR 1900+14][]{hcm+99},
and December 27, 2004 \citep[SGR 1806$-$20; ][]{hbs+05,mgv+05,bzh+07}.  These had peak
X-ray luminosities in the range $10^{44} - 10^{47}$~erg~s$^{-1}$, and each
was characterized by total energy release of over $10^{44}$~erg~s$^{-1}$
in the X-ray and soft-gamma-ray band.  The third flare was roughly 100
times more energetic than the first two and was by far the most luminous
transient yet observed in the Galaxy; it briefly outshone all the stars
in our Galaxy by a factor of 1000.  All three GFs have come from sources
traditionally called ``SGRs,'' and this one phenomenon appears to be the
{\it only} behavior that could in principle distinguish SGRs from AXPs;
on the other hand, SGR 0526$-$66 has now been inactive for several decades
since its GF \citep{kkm+03,tem+09}, and if it had been discovered during
that interval, probably would have been classified as an AXP.

\begin{figure}
\begin{minipage}{2.9in}
\hspace{-0.3in}
\vspace{0.2in}
\includegraphics[height=3in]{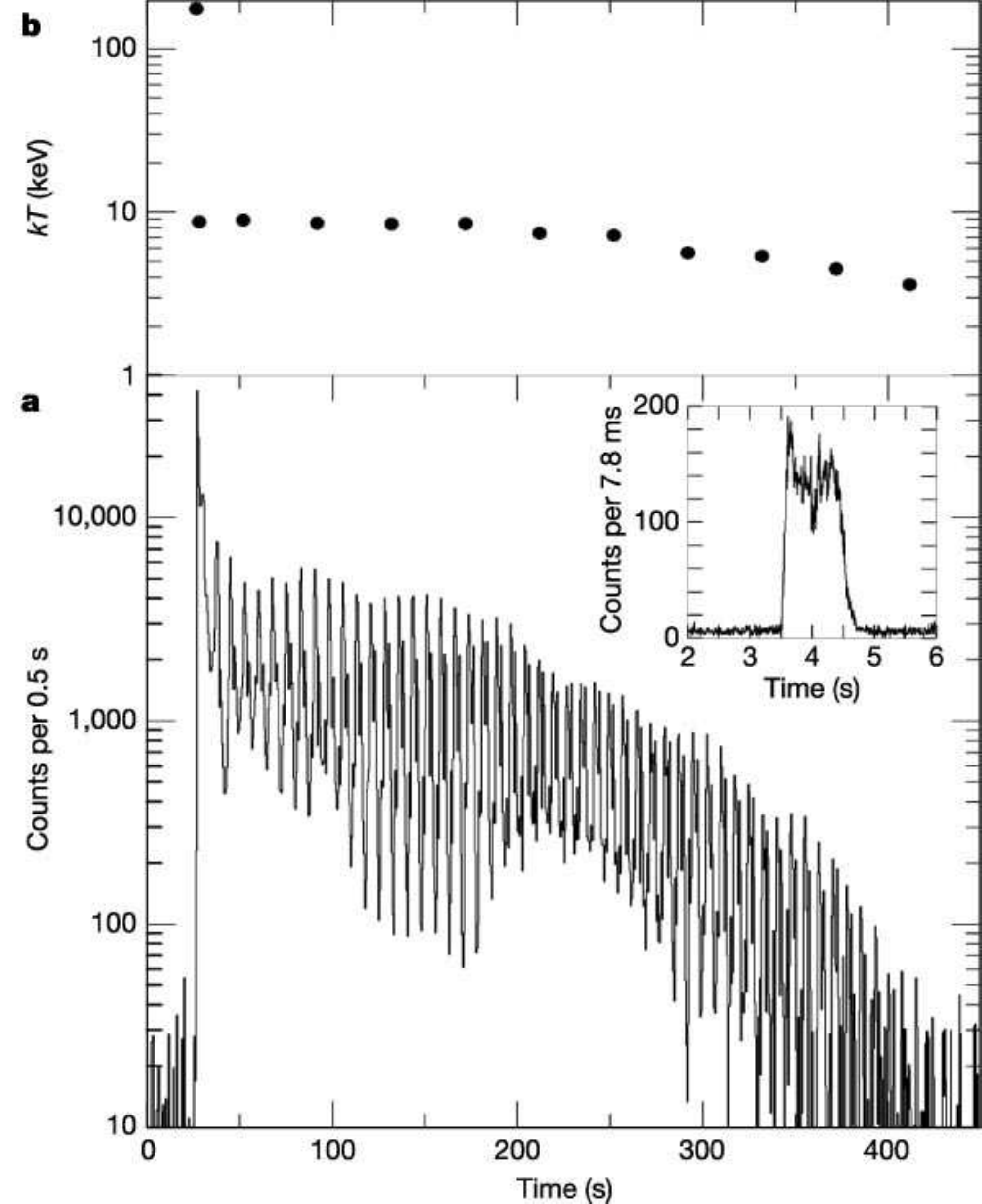}
\end{minipage}
\hfill
\begin{minipage}{2.9in}
\vspace{0.4in}
\hspace{-0.35in}
\includegraphics[scale=0.15]{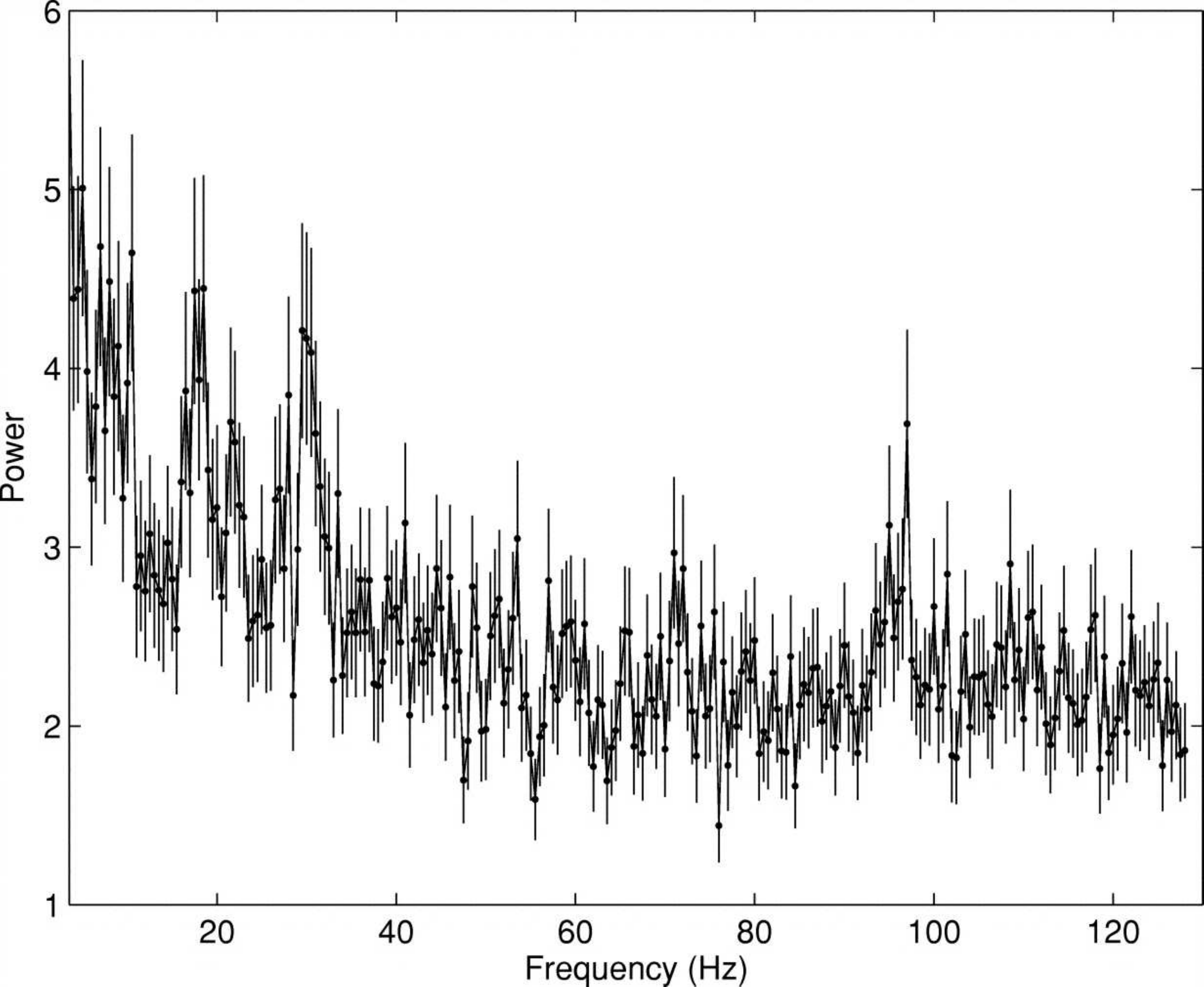}
\end{minipage}
\caption{(Left) The 2004 giant flare from SGR 1806$-$20.  a. 20--100-keV time history with 0.5-s resolution
from {\it RHESSI}, showing the initial spike (which saturated the detector) at 26 s.  The inset shows
a pre-cursor burst that occured just prior (with 8-ms resolution).  The oscillations in the decaying tail are at the neutron-star rotation period.  b. Blackbody temperature of the emission; see original
reference for details.  From \citet{hbs+05}.
(Right) Power spectrum of the SGR 1806$-$20 giant flare 2--80-keV light curve in the interval 200--300 s. 
Two low-frequency peaks at $\sim$18 and $\sim$30 Hz are visible, together with a small excess at $\sim$95 Hz.
From \citet{ibs+05}.
}
\label{fig:1806gf}
\end{figure}

The light curve and effective temperature evolution of the SGR
1806$-$20 giant flare is shown in Figure~\ref{fig:1806gf} (left) as an example
\citep[see][for details]{hbs+05}.
In the initial hard spike, lasting only $\sim$0.2~s,
over $10^{46}$~erg were released (assuming
a distance of 15 kpc).  Its peak luminosity was $10^{47}$~erg~s$^{-1}$.
This spike was followed by a several-minute long decay superimposed on
which are pulsations at the 7.56-s period of this source.  The total
fluence in this six-minute tail was $10^{44}$~erg.

%Giant flares are
%thought to be a result of a catastrophic instability in the star resulting in the
%eruption of a hot fireball consisting of a pair plasma, with little baryon
%contamination \citep{td95}.  The decaying portion of the giant flare is thought
%to represent the abatement of a trapped fireball, which is contained magnetically
%following the sudden initial outflow.  The pulsations are due to the rotation
%of the neutron star, which modulates the emission due to the magnetospheric
%emission anisotropy.  

The enormous peak luminosities observed in the giant flares together with their spectral
peak in the soft gamma-ray band makes them interesting possible counterparts of short, hard 
gamma-ray burst (GRB).  \citet{hbs+05} estimated that $\sim$40\% of all short, hard GRBs
detected by the BATSE instrument aboard the {\it Compton Gamma-Ray Observatory} could have
been GFs from distant extragalactic magnetars.  \citet{omq+08} suggested GRB 070201 may
have been a GF from a magnetar in M31, and \citet{hrb+10} suggested another such example,
though this interpretation is difficult to confirm in either case.  Moreover, young magnetars
cannot dominate the short GRB population as the latter are known to be commonly found at
large offsets from late-type galaxies \citep{ber14}.

\subsubsection{Quasi-Periodic Oscillations}
\label{sec:qpos}

A remarkable phenomenon detected during the pulsating tails of giant flares is the high-frequency
quasi-periodic oscillations (QPOs).
These are believed to be seismic vibrations
of the neuton star and may inform us on properties of the stellar interior.  
%Specifically, they may provide
%information on the internal magnetic field strength, the coupling between the crust and core, and on
%the dense matter equation of state.
These oscillations were first reported in phase-resolved portions of the tail emission of SGR~1806$-$20
following its 2004 GF \citep{ibs+05} at 92.5~Hz (see Fig.~\ref{fig:1806gf} right), 
and in phase-resolved emission also immediately post-GF in
the 1998 event from SGR~1900+14 at 84~Hz \citep{sw05}.  Several other QPO frequencies were also reported in these data,
which were obtained with {\it RXTE}.  Similar signals from SGR 1900+14 were also seen in {\it RHESSI} data
and new frequencies including a strong 626.5-Hz QPO, again only in specific rotational phase intervals but at a different time
\citep{ws06}.
%Some of these QPOs had extremely high amplitudes, in the range XXXX.
%LIFETIMES.
More recently, possible QPOs have also been reported from much fainter bursts from 1E 1547.0$-$5048
\citep{hdw+14}, however these oscillations have proven elusive in other sources
\citep[e.g.][]{hwu+13}.  
%The precise frequencies observed and their durations in some cases are consistent with expectations
%for stellar interior models \citep[e.g.][]{lev07}, but not always \citep[e.g.][ and see \S\ref{sec:andrei}]{gcs+12}.  
%Why certain frequencies appear but
%not others is also challenging to explain.  
%Sorely needed are more data containing QPOs.  However,
%as QPOs are clearly best detected in GFs, and GFs are very rare, the opportunities
%for understanding the origin of magnetar QPOs are few and far between and must await the next GF.

\subsection{Temporal Properties of Low-Frequency Emission}
\label{sec:multitemporal}

While every known magnetar has been detected as an X-ray pulsar, small handfuls have had
pulsations detected in the optical and radio bands.

Optical pulsations have been detected in just three magnetars \citep{km02,dmh+05,dml+09,dml+11}. 
%1E~1048.1$-$5937 \citep{dml+09} and SGR 0501+4516 \citep{dml+11}. 
The optical pulses seen thus far are comparably broad and similar to the corresponding X-ray pulse
profile, at least within the limited optical statistics. 
Importantly, all are detected with high pulsed
fractions ranging from 20\% to 50\%, in one case {\it higher} than in X-rays \citep{dml+11}. 
%which is hard to reconcile with the optical emission
%coming from reprocessing of the X-rays in a surrounding disk.  
This is strongly suggestive of a magnetospheric origin (see \S\ref{sec:andrei}).
Only three have had optical pulsations detected, and six more have shown optical and/or infrared emission not
yet seen to pulse (Table~\ref{ta:srcs}).  
%and Table 2 of the online catalog.  
%Near-infrared emission in particular (K band) is a favorite magnetar observing window, because of the high
%interstellar extinctions associated with magnetars' Galactic plane locations.
The overall picture of the relationship of the infrared emission to the X-rays remains unclear.
In several cases, clear infrared enhancements have been noted at the time of outbursts, often
with the infrared correlated with the X-rays \citep{tkvd04,icm+05,rtv+04}.
However in some cases, no correlation has been seen \citep{dk06,trm+08,wbk+08,tgd+08}.
%A clear correlation was seen between the infrared and X-ray fluxes after the 2002 outburst
%of 1E~2259+586, with identical flux decay indexes \citep{kgw+03,tkvd04}.
%\citet{icm+05} also reported similarity in the flux variability in X-rays and infrared emission.
%\citet{rtv+04} reported a correlation between infrared and X-ray fluxes following the outburst
%of XTE J1810$-$197, however \citet{crp+07} disputed this, and \citet{trm+08} later confirmed uncorrelated
%infrared variability.
%\citet{dk06} found a lack of correlation between variability in the infrared and the X-rays for 4U~0142+61.
%Near-infrared observations of 1E~1048.1$-$5937 suggest a possible correlation with X-ray flux,
%although the correlation does not seem exact \citep{wbk+08,tgd+08}.
%A discussion of the overall magnetar spectrum is deferred to \S\ref{sec:spectral}.

%\subsubsection{Radio}
%\label{sec:temporal_radio}

Radio pulsations have now been
detected in four magnetars \citep{crh+06,crhr07,lbb+10,sj13,efk+13},
%, 1E~1547.0$-$5408 \citep{crhr07},
%PSR J1622$-$4950 \citep{lbb+10}, and SGR J1745$-$2900 \citep{sj13,efk+13} 
five if counting the recent magnetar-like outburst from the 
radio pulsar PSR J1119$-$6127 \citep{akts16}.  
All four are `transient' magnetars and their radio emission
is transient too, associated with an X-ray outburst.
In XTE J1810$-$197 (the first detected radio-emitting magnetar), pulsations were
observed after, but not before, its 2003 outburst.  The radio pulsations
disappeared in late 2008, with no prior secular decrease in radio flux \citep{crh+16}.
Behavior in the other sources is similar \citep{crhr07,lbb+12}.
The persistent (i.e. non-transient) magnetars and several other transient magnetars 
have been searched for radio pulsations but with no detections 
\citep[e.g.][]{lkc+12,tyl13}.  
%Note that tentative claims of low-radio-frequency
%detections of some magnetars \citep[e.g.][]{mmt+05} have not to our knowledge been confirmed.
In the 4 magnetars with confirmed radio pulsations, the radio emission in all
cases is very bright, shows large pulse-to-pulse variability, with pulse morphologies,
both single and average, that can be broad and which generally change enormously on time scales of minutes.
They can be punctuated by spiky peaks that can be much shorter than the
pulse period.  This radio emission is also
highly linearly polarized, with polarization fractions of 60\%--100\%.
%The radio spectrum is discussed in \S\ref{sec:spectra_radio}.
No evidence for a radio burst at the epoch of a giant flare has been
seen \citep{tkp16}.

One radio-detected magnetar, 
SGR 1745$-$2900, is located in the Galactic Center \citep{mgz+13,kbk+13,sj13,efk+13}.
The radio properties of the magnetar are similar to those of the first three, 
with evidence for independent X-ray/radio flux evolution \citep{laks15,tek+15}. 
Notable is the far lower than expected interstellar scattering
\citep{sle+14,bdd+14}.  
%Although not relevant to our understanding of magnetars, 
This is exciting as it suggests renewed hope for finding
radio pulsars for dynamical studies near Sgr A* and for constraining properties
of the Sgr A* accretion disk \citep{efk+13}.
%Also, the highly linearly radio polarized pulse profile of SGR J1745$-$2900 also enabled
%\citet{efk+13} to constrain for the first time the strength of the magnetic field in the 
%accretion disk surrounding Sgr A*.

\section{RADIATION SPECTRUM}
\label{sec:spectral}

\subsection{Burst Spectra}
\label{sec:burstspectra}

\subsubsection{Short Bursts}

Bursts are generally spectrally much harder than the persistent X-ray
emission from magnetars and though easily detectable below 10~keV,
peak above that energy.  Hence, they are best studied by broadband
X-ray instruments or by combining simultaneous data from multiple
instruments, if possible.  Multiple different models have been used
to describe burst spectra, including a simple blackbody, double
blackbodies, optically thin thermal bremsstrahlung (OTTB) models, or
Componization models (a power law with an expontial cutoff).  Even with
very broadband spectra (e.g. 8--200 keV) it is hard to distinguish among
these models \citep[e.g.][]{vkg+12}; moreover
sometimes the overall spectra of bursts in a cluster changes with epoch
\citep[e.g.][]{vgk+12}.  Regardless of model, in all studies, spectral
hardness is found to be related to burst fluence.  In most cases the
two appear anti-correlated \citep{gwk+99,gwk+00,gkw+01,vkg+12} and in
some cases correlated \citep{gkw04}.  Tail spectra are typically well
modelled by blackbodies that show decreasing $kT$ at constant radius
\citep[e.g.][]{gwk+11,akb+14}.

%\subsubsection{Spectral Features in Short Bursts}

Curiously, an apparent emission feature near
$\sim$13~keV has been noted in the spectra of bright magnetar bursts detected with {\it
RXTE}/PCA in the sources 1E~1048.1$-$5937, 4U~0142+61 and XTE~J1810$-$197
\citep{gkw02,wkg+05,gdk11,cgsk16}.  These features are variable,
occuring transiently during the burst evolution, but are not subtle:
they are easily visible by eye, with equivalent widths of $\sim$1~keV.
Although it is tempting to argue these features are of some unknown
instrumental origin (even though they are not always seen in {\it RXTE}
magnetar burst data), \citet{akb+14} found evidence of it in {\it NuSTAR}
data from 1E~1048.1$-$5937, which happened to burst during an observation.
The origin of these burst features is unknown; if some form of cyclotron
emission, it is unclear why all the sources in which it has been observed
show it near the same energy, since presumably they have a variety of
field strengths in the emission region. There is no spectral line known
with that energy.

\subsubsection{Giant Flares}

Described in \S\ref{sec:giantflares}, the initial brief spike seen in magnetar
giant flares is extremely hard, peaking in the soft gamma-ray band and extending
at least to MeV energies. It is
followed by a softening on time scales of seconds to minutes. In
Figure~\ref{fig:1806gf}, where the 2004 giant flare of SGR 1806$-$20 is shown, a blackbody
describes the data reasonably well, and
the initial $kT$ is $\sim$175~keV \citep{hbs+05,bzh+07}.  
Following the spike was a few-second decay whose spectrum was non-thermal, described
by a power law of index $\Gamma \sim 1.4$ and then a series of pulsations at the 7.5-s spin period, with spectra consisting
of a combination of blackbody and power-law emission.  Both softened over the next few minutes,
from $kT \simeq 11$~keV to 3.5~keV and $\Gamma \simeq 1.7$ to 2, for the two components,
respectively.
Similar initial and subsequent spectra were seen in the
1998 giant flare from SGR 1900+14 \citep{hcm+99}.  Although at the peak of the
1979 giant flare from SGR 0526$-$66 the spectrum may have been slightly softer than in the latter
two cases (the best estimate is $\sim$30~keV), 
the overall trend of a very hard spike and subsequent softening were also seen \citep{fek+81}. 
In some ways, giant flare spectral evolution mirrors that of the X-ray emission in less-energetic
magnetar outbursts:  sudden hardening and subsequent softening; see \S\ref{sec:specevol}.  
%However, the emission in giant flares is thought to originate from an expanding e$^+$/e$^-$ fireball
%triggered by a large-scale reconnection/interchange instability of the stellar magnetic field \citep{td95},
%rather than by a simple magnetospheric twist.  This is discussed in more detail in \S\ref{sec:andrei}.

%However in giant flares,
%due to instrumental saturation, the true spectrum of the initial spike is not
%known with certainty; nevertheless the total flux at Earth has been estimated to be
%1 erg~cm$^{-2}$!

\subsection{Persistent Emission}
\label{sec:persistentspectra}

The X-ray spectra of magnetars in quiescence fall into two broad classes:
those in high-quiescent-luminosity sources (the `persistent' magnetars)
and those in low-quiescent-luminosity sources (the `transient' magnetars).
The observational distinction is based both on quiescent luminosity ($\gapp 10^{33}$~erg~s$^{-1}$
for persistent sources like 1E 2259+586 or 4U 0142+61 versus
$\lapp 10^{33}$~erg~s$^{-1}$ for transient sources like XTE J1810$-$197 or
SGR J1745$-$2900) and on flux dynamic range in outbursts (factor of $\lapp$100 in persistent
sources versus $\gapp$100 in transient sources).

\begin{figure}
\begin{minipage}{2.9in}
\includegraphics[scale=0.3]{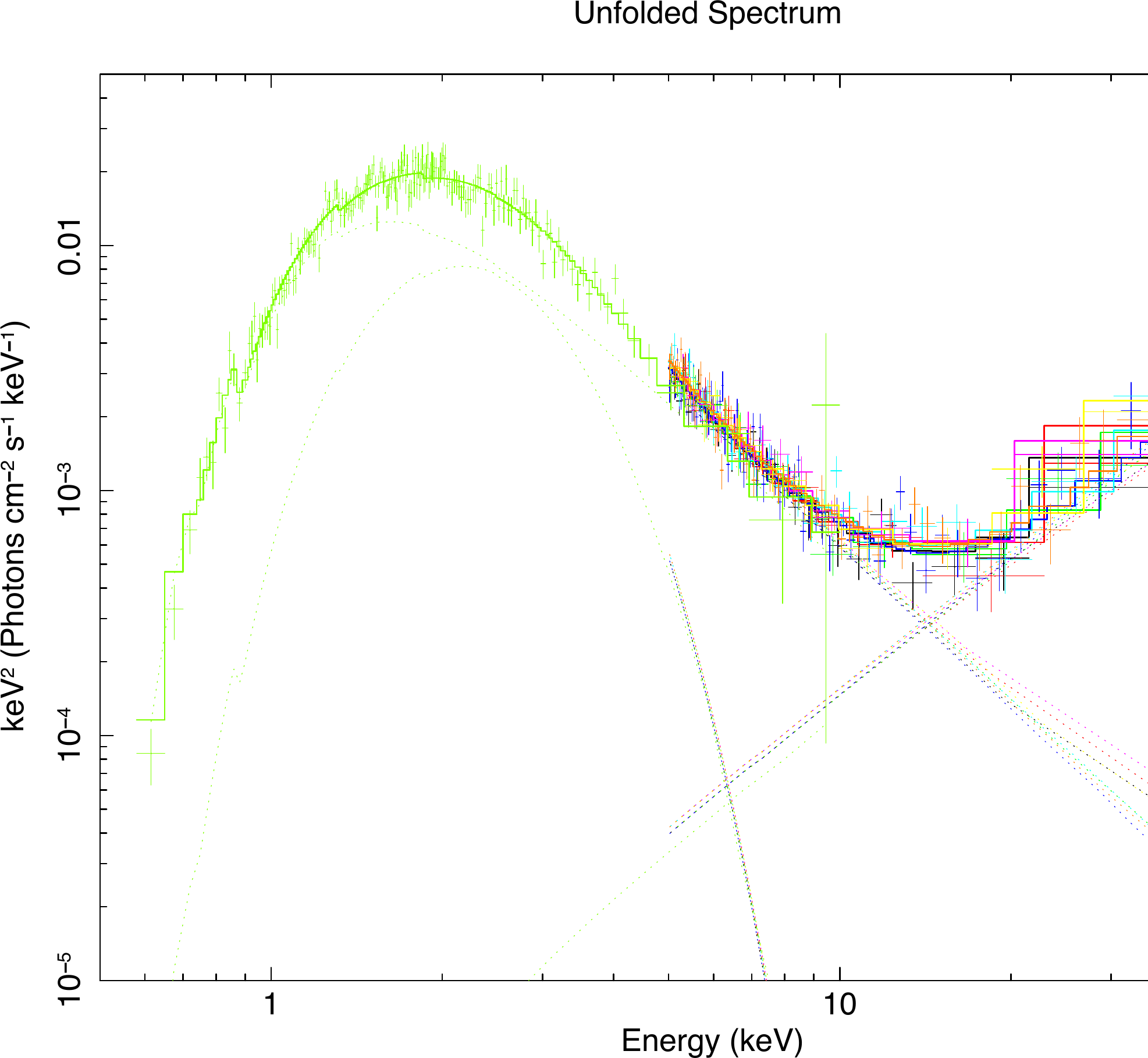}
\end{minipage}
\hfill
\hspace{-0.3in}
\begin{minipage}{2.9in}
\includegraphics[scale=0.22]{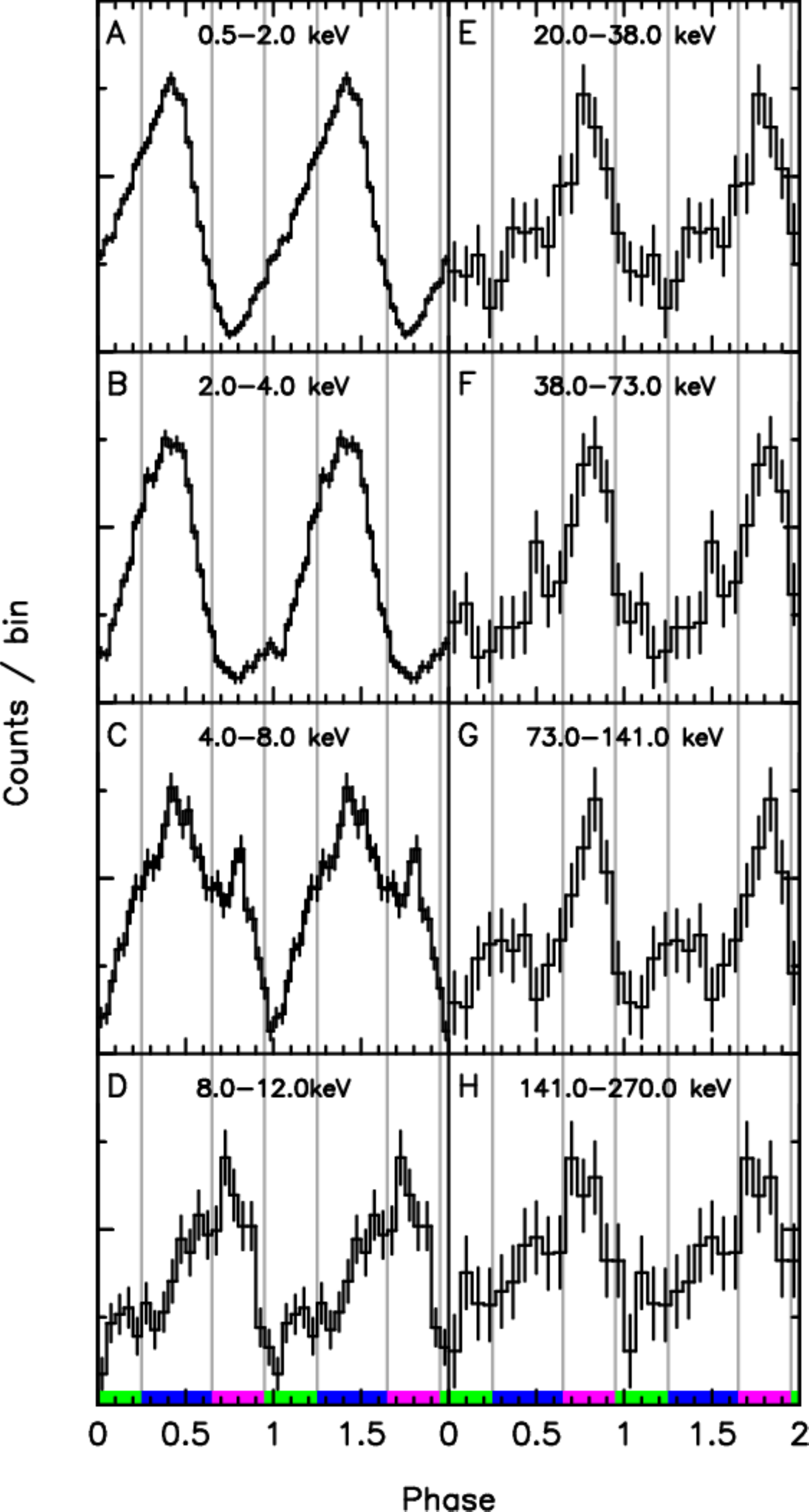}
\end{minipage}
\caption{
(Left) 
Broadband phase-averaged X-ray spectrum from combined {\it Swift}/XRT (green) and {\it NuSTAR} observations of 1E~2259+586
from \citet{vhk+14}.  The best-fit model of an absorbed blackbody plus two power laws is shown.  The spectral
turn up in this source near 15 keV is obvious.
(Right) 
Pulse profiles in different X-ray energy bands for 1RXS J170849-400910,
from \citet{dkh08}.}
\label{fig:spectrum}
\end{figure}

\noindent
{\it Classic Magnetars in quiescence}
show multiple-component X-ray spectra that are usually well parameterized in the 0.5--10-keV band
by an absorbed blackbody of $kT \simeq 0.3-0.5$~keV plus a power-law component of photon
index in the range $-2$ to $-4$ (see OK14 for a compilation).  An example of such
a spectrum is shown in Figure~\ref{fig:spectrum} (left).   Typically
the non-thermal component begins to dominate the spectrum above $\sim$3--4 keV.
At least qualitatively, the thermal component is thought to arise from the hot neutron-star
surface, while the power-law tail likely arises from a combination of atmospheric 
and magnetospheric effects.
%A correlation between power-law index and spin-down rate was reported by \citep{mw01} and
%later confirmed by \citet{kb10} and is a cornerstone of twisted magnetosphere models
%(see \S\ref{sec:andrei} below).
%In these classic spectra, a diversity in the relative normalizations 
%of the two components is seen.  
Typically the power law dominates energetically by
at least a factor of two, although in the soft X-ray band this quantity depends
strongly on the absorption, since $N_H$ and $kT$ are generally highly covariant. 
Double blackbodies can often fit magnetar spectra as well, and in rare cases double power laws.
Note, however, as described in \S\ref{sec:andrei}, the soft-band X-ray spectrum is thought
to arise physically from a complicated blending of surface thermal emission distorted
by the presence of a highly magnetized atmosphere, then Comptonized by currents
in the magnetosphere which can themselves result in surface heating via return currents.
Hence the simple e.g. blackbody plus power-law or double-blackbody parameterizations should
be seen as merely convenient and readily available (i.e. in XSPEC) descriptions of the data 
rather than actual measurements of physical properties.  
%The latter requires detailed fitting
%of custom models that include all the above effects, as described in \S\ref{sec:andrei}. 

Observations using {\it INTEGRAL} and the {\it Rossi X-ray Timing Explorer}
led to the surprising discovery just over a decade ago
that for persistent magnetars, the spectrum turns up above 10 keV, such that the bulk of
their energy comes out {\it above} the traditional 0.5--10-keV band \citep{khm04,khdc06}.
This prominent hard spectral component is shown for magnetar 1E 2259+586 in Figure~\ref{fig:spectrum} (left).
Such hard components have been seen now in six sources in quiescence.
\citet{kb10} and \citet{enk+10} 
reported a possible anti-correlation between degree of spectral up-turn and spin-down rate 
and/or spin-inferred magnetic field strength,
such that higher spin-inferred-B sources show little to no
spectral up-turn \citep{mgmh05,gmte06}.
%1048 an exception \cite{turkey,wang}

Another remarkable feature of magnetar spectra is that they are highly
rotational-phase dependent \citep{dkh08,dkh+08}.  This is diagnosed in two different ways:  by
strong energy-dependence of the pulse profile as shown for 
magnetar 1RXS J170849$-$400910 in Figure~\ref{fig:spectrum} (right),
or, equivalently, as variations in fitted spectral parameters with rotational phase.
The latter is detected generically.
%, at least in sources for which there
%is sufficient flux to allow a meaningful fit in many phase bins.  
The strong phase variation is expected for magnetospheric emission beamed along magnetic field lines
and has been used to deduce constraints on the geometry of the hard X-ray source
\citep{hbd14,vhk+14,aah+15,thy+15}.
This is discussed in detail in \S\ref{sec:andrei}. 

\noindent
{\it Transient Magnetars in quiescence}
show X-ray spectra that are consistent with being
pure absorbed blackbodies, with $kT \simeq 0.15-0.3$~keV (see OK14 for a compilation).
As these sources are generally
discovered in outburst, measuring quiescent spectra and temperatures requires waiting
months to years until the source returns to a quiescent state \citep[e.g.][]{ah16}.
Phase-resolved spectroscopy for transient magnetars in quiescence has yet to be done owing to the faintness
of the sources.  In these objects, the non-thermal magnetosperic processes responsible
for the power laws seen in persistent magnetars seem absent, in spite of the spin properties of
the two classes being similar, though sensitivity may play a role.  
Interestingly, quiescent transient magnetar spectra are similar to the
X-ray spectra of some high-B radio pulsars (see \S\ref{sec:highB}).
%notably those with sufficiently low spin-down luminosity
%that non-thermal
%emission is not a contaminant.  In the latter sources, the X-ray spectra
%appear to be purely thermal, with blackbody $kT \simeq 0.15-0.2$~keV.
%Interestingly, there is evidence that their temperatures are higher than
%in radio pulsars of similar age but lower field \citep{km05,zkm+11,ozv+13}
%suggesting possible internal heating due to the high field.  Also,
%\citet{akt+12} showed evidence for a correlation between quiescent X-ray
%luminosity and spin-inferred magnetic field strength in high-B radio
%pulsars and magnetars, with the two source classes lying on the same
%correlation curve.  These observations suggest that high-B radio pulsars
%could be quiescent magnetars, a prediction validated by magnetar-like
%outbursts from two rotation-powered pulsars (see \S\ref{sec:highB}).
%Conversely, they also suggest that transient magnetars might show
%conventional radio pulsations in quiescence, a prediction that has not
%thus far been verified, though additional sensitive searches during
%quiescence are needed, and radio beams are narrow so that the chance of
%detecting any one source is likely limited.

\subsection{X-ray Spectral Evolution in Outburst}
\label{sec:specevol}

\begin{figure}
%\begin{minipage}{1.2in}
%\hspace{-2.0in}
%\includegraphics[scale=0.13]{2259}
%\end{minipage}
%\hfill
%\hspace{0.45in}
%\begin{minipage}{3.0in}
\includegraphics[scale=0.37]{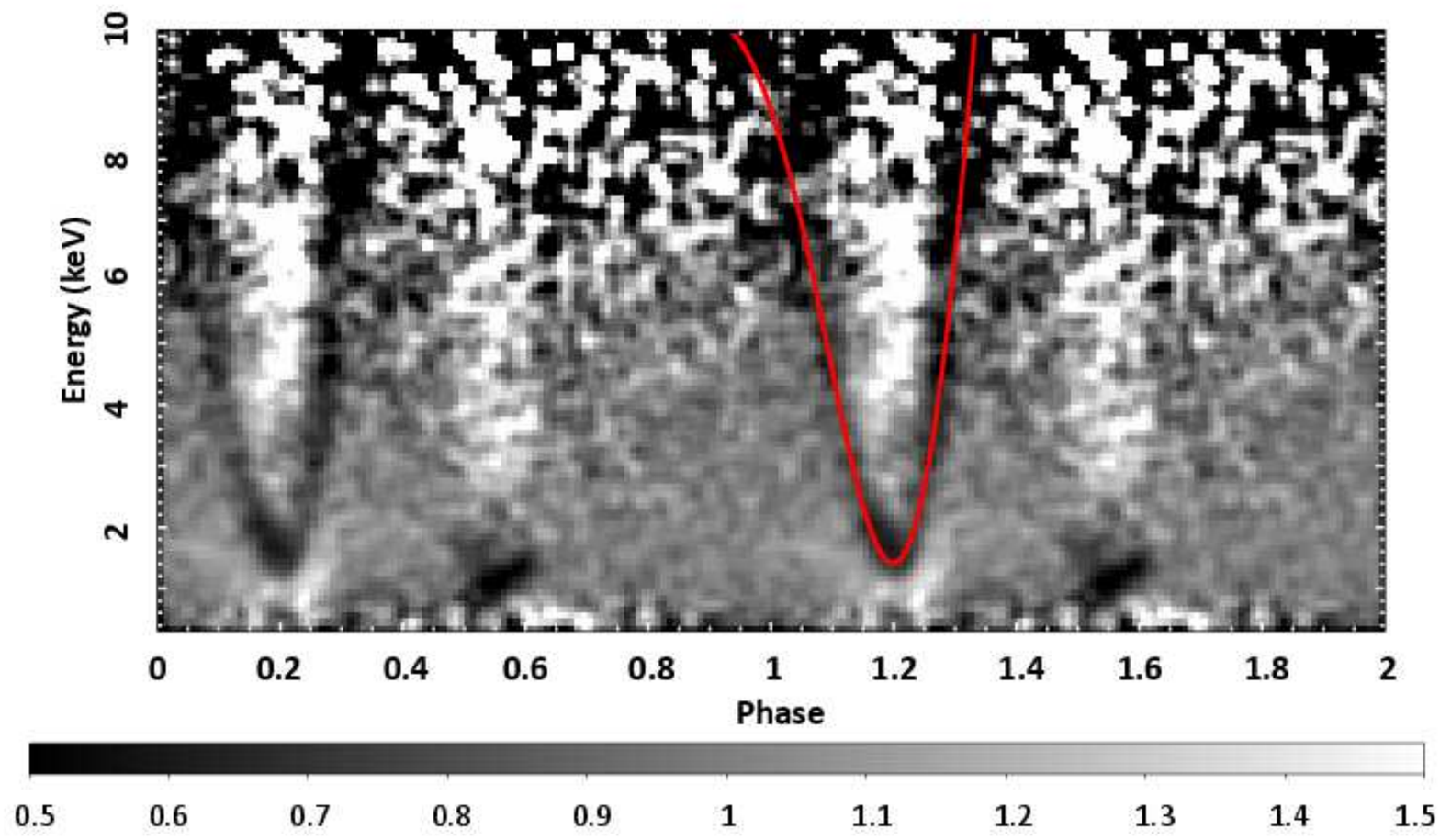}
%\end{minipage}
\caption{
Phase-resolved spectroscopy of SGR 0418+5729.
The spectral flux is shown in the energy versus phase plane for {\it XMM-Newton} EPIC data, with
100 phase bins and 100-eV energy channels.
The red line shows (for only one of the two displayed cycles) 
a simple proton cyclotron model. 
See \citet{tem+13} for details.}
\label{fig:specfeature}
\end{figure}

The spectra of magnetars change dramatically at times of outburst,
generically hardening initially, then slowly softening as the flux
relaxes back to quiescence over typically months to years.  The flux
evolution in outbursts was discussed in \S\ref{sec:outbursts}, with an
example shown in Figure~\ref{fig:outburst} (left).
The hardening at outburst, for a spectrum parameterized by an absorbed
blackbody plus power law, can generally be described by an initial increase
in $kT$ by a factor of $\sim$2--3 (often, but not always, with a decrease in effective blackbody
radius by a factor of a few), together with a decrease in photon index
by a factor of $\sim$2.  These quantities then relax back to their
quiescent values on the same time scale as the flux relaxation.  
An example of the spectral evolution seen in one magnetar (1E 2259+586)
outburst is shown in Figure~\ref{fig:specfeature}.
The hardness/flux correlation seen in magnetar outbursts is thought to
be closely related to the correlation between hardness and spin-down rate
noted by \citet{mw01}, in that all these quantities
are related to the degree of magnetspheric twist, with larger twists corresponding
to higher luminosities, spin-down rates, and hardness.  

There is, however, considerable diversity in the spectral changes and evolution
post-outburst.  Just as relaxation light curves for different sources can
look very different, the degree of hardening
and the manifestation of that hardening (be it a greater increase in $kT$ or decrease
in photon index) varies from outburst to outburst \citep{re11}.  
In the first-discovered transient magnetar, XTE J1810$-$197, for example,
\citet{gh07} found that the spectrum was well described by two blackbodies each
of which had its luminosity decay on exponential time scales, albeit different ones (870 and 280 days),
behavior not reproduced in most other sources.
Even for the same source, outbursts can show a variety
of behaviors \citep{ier+10,nkd+11,khdu12}.
\citet{sk11} show that there does not appear to exist a universal law linking the degree
of flux increase over the quiescent level with the degree of flux
hardening.   On the other hand, \citet{bl16} showed a relationship
between X-ray luminosity and inferred blackbody emitting area during outburst relaxations
of 7 different magnetars, consistent with theoretical predictions based on j-bundle
untwisting (see \S\ref{sec:andrei}).

%\begin{figure}
%\includegraphics[scale=0.15]{2259}
%\caption{Spectral evolution of 1E 2259+586 through and following its 2002 outburst. 
%Top to bottom: unabsorbed flux (2–10 keV), blackbody temperature (kT), 
%photon index, blackbody radius, and ratio of power-law (2–10 keV) to bolometric blackbody flux. 
%Horizontal dashed lines denote the values of each parameter fortuitously measured 
%one week prior to the outburst.  From \citet{wkt+04}.}
%\label{fig:specevol}
%\end{figure}

\subsubsection{Spectral Features}

\citet{tem+13} reported the presence of a feature -- an absorption line --
in the outburst X-ray spectrum of SGR 0418+5729, the source
with low spin-inferred magnetic field \citep{ret+10}.  The
energy of the line apparently varies strongly with pulse phase;
see Figure~\ref{fig:specfeature}.  The variation in energy is roughly a factor of 5
over just 10\% of the pulse phase.
Those authors interpret the line as a proton cyclotron feature; its energy implies a magnetic field
ranging from $2 \times 10^{14}$~G to $>10^{15}$~G.  If interpreted as an electron cyclotron line, however,
the implied field is 2000$\times$ lower.  If the proton cyclotron interpretation is correct,
this observation strongly supports the hypothesis that SGR 0418+5729 has a far stronger field than
is inferred from the dipolar component. 
A similar phase-dependent absorption line was
recently reported by \citet{rit+16} for the magnetar Swift J1822.3$-$1606.  This is particularly
interesting as this source has the second lowest spin-inferred B field of the known magnetars 
\citep{rie+12,skc14,rit+16}.  Why the two lowest-inferred B sources
should be the only ones with such phase-dependent features is unclear.  The emission may come
from a magnetic loop near the surface of the star, wherein the field
energy is appropriate for the line to be in the observed spectral window.
%In higher-inferred-field magnetars, perhaps the loops have
%much higher B, so that the lines are in the fainter higher-energy portion of the X-ray spectrum
%where the source is fainter.  In this case, more sensitive hard X-ray telescopes may be able
%to see these features in more magnetars.

\subsection{Low-Frequency Emission}
\label{sec:spectra_oir}

%The origin and nature of optical and infrared (IR) emission in magnetars is not well understood.
As discussed in \S\ref{sec:multitemporal}, six magnetars have
optical or IR emission detected.
One is bright enough to have had its optical/IR spectrum studied
in detail:  4U~0142+61 \citep{wck06}.  
%Its broadband optical/IR spectrum from
%\citet{wck06} is shown in Figure~\ref{fig:oir_spectrum}.  
The optical/IR 
emission is well described by power law of index 0.3 and is presumed to be magnetospheric,
in line with the detection of strong optical pulsations \citep{km02,dmh+05}.
However, the IR {\it Spitzer}-measured 4.5 and 8.0 $\mu$m emission deviates from this function
and can be well described by blackbody emission for a temperature of 920 K.  
\citet{wck06} interpreted this near-IR emission as arising from an X-ray heated dust disk that is a remnant
of material that fell back toward the newly born neutron star following the supernova.
%Here `passive' means that the disk does not provide the source's X-ray luminosity
%as in accreting neutron stars, but rather is heated by the magnetar's X-ray emission so that
%it shines in the infrared range.  They estimate a disk of mass $\sim$10 Earth masses.  
However, this interpretation is not unique; 
it may also be some form of non-thermal magnetospheric emission.
The detection of pulsations in the near-IR would be key as 
these are not expected at more than the few-percent level for a disk.  
%However, such an observation has yet to be done.  
Note that \citet{wbk+08} did deep {\it Spitzer} observations of magnetar 1E~1048.1$-$5937 that
appear to rule out any IR emission similar to that in 4U~0142+61, and hence challenge
the disk interpretation.

%\begin{figure}
%\includegraphics[height=2.5in]{oir_spectrum}
%\caption{ Optical/infrared spectral energy distribution
%of 4U 0142+61 from \citet{wck06}.
%The triangles indicate the observed optical/IR
%flux while the squares indicate the dereddened flux assuming
%AV = 3.5. The optical emission is assumed here to be magnetospheric,
%and is well described by a power law of index 0.3.  The 4.5 and
%8.0 $\mu$m emission, however, does not follow this power law, but
%is well described by a blackbody of temperature 920 K, as shown.
%}
%\label{fig:oir_spectrum}
%\end{figure}

%\subsubsection{Radio Emission}
%\label{sec:spectra_radio}

The spectrum of the pulsating radio emission seen at least transiently from
four magnetars is remarkably flat \citep{crp+07,crhr07,ljk+08,lbb+10}. 
This is true even over a very wide range of radio frequencies
\citep[e.g. 1.4--45 GHz for 1E~1547.0$-$5408][]{crj+08}.  
This flatness is in contrast to the spectra of rotation-powered pulsars which are typically
steep, with negative spectral indexes of $\sim -1.8$ \citep[e.g.][]{mkkw00a}.
%Indeed for an extended period after its 2003 outburst, XTE J1810$-$197 had
%greater flux above 20 GHz than any known radio pulsar \citep{crh+06}.
Magnetar SGR J1745$-$2900
has been detected at frequencies as high as 225 GHz, the highest
yet for any pulsar \citep{tek+15}.  On the other hand, for this same source,
\citet{ppe+15} reported a steep spectral index ($-$1.4) between 2 and 9 GHz.
Although overall approximately flat, magnetar radio spectra may not be well described by a single spectral index.
%That said, the actual functional form of the spectrum is not well known, as 
Measurement of the broadband radio spectra is 
challenging because of the high variability and the common presence of
terrestrial interference at relevant time scales.

%The relationship of the radio spectrum to that of the X-rays is
%not well understood.   While it is clear that radio emission is broadly correlated with the X-rays
%in that the former appears near X-ray outbursts, the fading time scales are not obviously the same,
%with the X-rays showing typically less short time scale variability during the post-outburst phase 
%but more secular, long-term decay.  For example, \citet{laks15}
%found that the 2–10-keV X-ray flux decayed steadily in 
%SGR J1745$-$2900 following the source's initial X-ray outburst, 
%while the average radio flux remained stable to within ∼20% during a 5.5-month-long stable state.
%However, even while the X-rays continued their slow fading, the radio emission susbsequently
%became more erratic.

\section{MECHANISM OF MAGNETAR ACTIVITY}
\label{sec:andrei}
% ar-sample-1col.tex, dated 30th Mar. 2013
% This is a sample file for AR journals
%
% Compilation using 'ar.cls' - version 1.0, Aptara Inc.
% (c) 2013 AR
%
% Steps to compile: latex latex latex
%
% For tracking purposes => this is v1.0 - Mar. 2013

%\documentclass{ar-1col}
%
%\usepackage{rotating}%
%\usepackage{subfigure}
%\usepackage{natbib}
%\usepackage{aas_macros}
%\bibliographystyle{ar-style2}
%
%
%\usepackage[numbers]{natbib}%
%
% \usepackage[T1]{fontenc}
% \usepackage[latin1]{inputenc}
% \usepackage{babel}
% \usepackage[font=small,labelfont=bf,tableposition=top]{caption}
% \usepackage[font=footnotesize]{subfig}
% \usepackage{blindtext}
% 
% Metadata Information
%\jname{Xxxx. Xxx. Xxx. Xxx.}
%\jvol{00}
%\jyear{YYYY}
%\doi{10.1146/((please add article doi))}

%%%%%%%%%%%%%%%%%%%%%%%%%%%%%%%%%%%%%%%%
% \def\simlt{\lower.5ex\hbox{$\; \buildrel < \over \sim \;$}}
% \def\simgt{\lower.5ex\hbox{$\; \buildrel > \over \sim \;$}}
\def\XL{\textcolor{red}}
\def\AB{\textcolor{black}}

\def\sT{\sigma_{\rm T}}
\def\dd{{\rm d}}

\def\be{\begin{equation}}
\def\ee{\end{equation}}
\def\beq{\begin{eqnarray}}
\def\eeq{\end{eqnarray}}
\def\bA{{\,\mathbf A}}
\def\bB{{\,\mathbf B}}
\def\bE{{\,\mathbf E}}
\def\bj{{\,\mathbf j}}
\def\bv{{\,\mathbf v}}
\def\jB{j_B}
\def\RNS{R}
\def\Rmax{R_{\rm max}}
\def\Rlc{R_{\rm lc}}

\def\fm{f}
\def\fR{f_R}
\def\ta{\psi}
\def\tamax{\psi_{\rm max}}
\def\Etw{E_{\rm tw}}
  \def\Etor{E_{\rm tw}}
\def\tf{t_{\rm twist}}
\def\tend{t_{\rm end}}
  \def\Iopen{I_{\rm lc}}
  \def\Lopen{L_{\rm lc}}
\def\uf{u_{\star}}
\def\vf{v_{\star}}
\def\uuu{\tilde{u}_{\rm twist}}
\def\BQ{B_Q}
\def\js{j_\star}
\def\V{{\cal V}}
\def\LV{L_{\cal V}}
\def\tV{t_{\cal V}}
\def\Vei{{\cal V}_{ei}}
\def\Vpm{{\cal V}_{\pm}}
\def\gres{\gamma_{\rm res}}
\def\bOm{{\bf\Omega}}
\def\tev{t_{\rm ev}}
\def\tg{t_{\rm delay}}
\def\VGeV{V_9}
\def\Lradio{L_{\rm radio}}
\def\Lobs{L_{\rm obs}}
\def\Aobs{A_{\rm obs}}
\def\epsradio{\epsilon_{\rm radio}}
\def\XTE{XTE~J1810-197}
\def\AXP{1E~1547.0-5408}
\def\sigcr{\sigma_{\rm cr}}
\def\scr{s_{\rm cr}}

\def\rhoGJ{\rho_{\rm GJ}}
\def\rpc{r_{\rm pc}}
\def\Epar{E_\parallel}
\def\br{{\mathbf r}}
\def\M{{\cal M}}
\def\nmin{n_{\rm min}}
\def\vH{v_{\rm H}}
\def\T{{\cal T}}
\def\Tc{T_{\rm core}}
\def\thB{\Theta_B}
\def\Tcrit{T_{\rm crit}}
\def\Lum{L}
\def\ff{f_\star}
\def\Phie{\Phi_\parallel}
\def\vA{v_{\rm A}}
\def\Emag{E_{\rm mag}}
\def\sSB{\sigma_{\rm SB}}
\def\Eq{Equation}
\def\RLC{R_{\rm LC}}
\def\Phipc{\Phi_0}
\def\fopen{f_{\rm open}}
\def\tohm{t_{\rm ohm}}
\def\kB{k_{\rm B}}
\def\bk{{\mathbf k}}

%%%%
\newbox\grsign \setbox\grsign=\hbox{$>$} \newdimen\grdimen \grdimen=\ht\grsign
\newbox\simlessbox \newbox\simgreatbox \newbox\simpropbox
\setbox\simgreatbox=\hbox{\raise.5ex\hbox{$>$}\llap
     {\lower.5ex\hbox{$\sim$}}}\ht1=\grdimen\dp1=0pt
\setbox\simlessbox=\hbox{\raise.5ex\hbox{$<$}\llap
     {\lower.5ex\hbox{$\sim$}}}\ht2=\grdimen\dp2=0pt
\setbox\simpropbox=\hbox{\raise.5ex\hbox{$\propto$}\llap
     {\lower.5ex\hbox{$\sim$}}}\ht2=\grdimen\dp2=0pt
\def\simgt{\mathrel{\copy\simgreatbox}}
\def\simlt{\mathrel{\copy\simlessbox}}
\def\simprop{\mathrel{\copy\simpropbox}}
%%%%

%%%%%%%%%%%%%%%%%%%%%%%%%%%%%%%%%%%%%%%%
% Document starts
%\begin{document}

% Page heads
%\markboth{Kaspi \& Beloborodov}{Magnetars}

% Title
%\title{Magnetars}

% Author/affiliation
%\author{Victoria M. Kaspi,$^1$ Andrei M. Beloborodov$^2$%
%\affil{$^1$Department of Physics, McGill University,
% Rutherford Physics Building, 3600 University Street, 
%Montreal, Quebec, H3A 2T8; 
%email: vkaspi@physicsl.mcgill.ca}
%\affil{$^2$Department of Physics, 
% and Columbia Astrophysics Laboratory,
%Columbia University, New York, New York 10027;
%email: amb@phys.columbia.edu}
% \affil{$^3$Department of Chemistry, Purdue University, West Lafayette, Indiana 47907; email: lslipchenko@purdue.edu}}

% First page note
%\firstpagenote{This is an example of dummy text used to illustrate an example of first page note.}

% Abstract
%\begin{abstract}
%\end{abstract}

% Keywords
%\begin{keywords}
% ultrafast, core electron correlation, coherence, stimulated Raman
%\end{keywords}

%\maketitle
%
% to generate article TOC
%\tableofcontents

%\section{MECHANISM OF MAGNETAR ACTIVITY}

\subsection{Internal Dynamics}

A nascent magnetar experiences fast evolution in the first minutes of its life.
Magnetohydrodynamic (MHD) relaxation leads to a magnetic configuration that has a 
strong toroidal component \citep{bra09}, 
whose stability is assisted by compositional stratification \citep{arm+13}.
Neutrino cooling leads to crystallization of the crust. The resulting object 
has a liquid core of radius $\sim 10$~km surrounded by a 1 km-thick solid crust. 

The 
% dynamic 
\AB{subsequent }
behavior of magnetars 
on the timescales of 1-10~kyr
is associated with slow evolution of the magnetic field inside the star, which is 
capable of breaking the solid crust.
The interior of a neutron star is an excellent conductor,
and hence the magnetic field is practically ``frozen'' in the stellar material. 
More precisely, the magnetic field is frozen in the electron fluid,
and field evolution is possible due to the multi-fluid composition of the star.
The electrons may slowly drift with respect to neutrons and also with respect to ions. 
This gives two processes capable of  moving the magnetic field lines: 
ambipolar diffusion and Hall drift \citep{gr92}.

\subsubsection{Ambipolar diffusion}

Ambipolar diffusion is the motion of the electron-proton plasma 
(coupled with the magnetic field) through the static neutron liquid in the core. 
The magnetar is born with  electric currents $\bj=(c/4\pi)\nabla\times\bB$, and
hence  Lorentz forces $\bj\times\bB/c$ are applied to the plasma.
The plasma does not move in a young hot magnetar -- it is 
stuck in the heavy neutron liquid due to frequent proton-neutron collisions.
As the star ages and cools down, the p-n collision rate decreases 
\citep{ys90}, and ambipolar drift develops on the timescale 
$t_{\rm amb}\sim 10^3 (T_9/k_{-5}B_{16})^2$~yr,\footnote{
   Hereafter we use the standard notation $X_m$ for a quantity $X$ normalized to
   $10^m$ in CGS units.} 
where the wavenumber
$k\sim 10^{-5}$~cm$^{-1}$ describes the gradient of the magnetic field in the core
and corresponds to a characteristic scale $\pi/k\simlt R$.
The timescale of ambipolar diffusion becomes comparable to the 
star's age when the core temperature 
decreases to $\sim 10^9$~K. Then a significant drift occurs
and a large fraction of magnetic energy can be dissipated by the p-n friction.
As the drift develops, it generates plasma pressure gradients, which 
tend to be erased by weak interactions $e+p\leftrightarrow n$ \citep{gr92,td96a}.
\AB{Core superfluidity is capable of suppressing these processes and quickly ending
ambipolar diffusion \citep{gjs11}. However, for a plausible critical temperature of superfluidity,
the suppression effect is moderate, and ambipolar diffusion can still be the main cause of 
magnetar activity \citep{bl16}.}
The ambipolar drift tends to relax the magnetic stresses that drive it and 
eventually stalls while the core temperature drops.

\subsubsection{Hall drift}

Hall drift is the transport of magnetic field lines by the electric current $\bj$,
which implies a flow of the electron fluid relative
to the ions with velocity ${\mathbf v}_{\rm H}=-\mathbf j/en_e$,
where $n_e$ is the electron density. 
 Hall drift is normally very slow in the core, because of its high density
$n_e\sim 10^{37}$~cm$^{-3}$, but can be significant in the crust. 

Hall drift conserves the total magnetic energy, however it can generate new electric 
currents. When ohmic dissipation is taken into account, the evolution 
may come to a steady state \citep{gc14}. The field evolution in the crust 
was simulated numerically for axisymmetric configurations (e.g. \cite{pmg09}) 
and in more general 3D configurations \citep{gwh16}.
It was seen to build up strong elastic stresses in the crust \citep{pp11}.
Hall waves can also be excited near the crust-core boundary \citep{td96a}. 
% These waves propagate outward and cause significant crustal deformations \citep{llb16}.

\subsubsection{Mechanical failures of the crust}

The ambipolar and Hall drift of the magnetic field lines results in the gradual accumulation 
of crustal stresses which can trigger surface motions in magnetars.
The crust is nearly incompressible, however its Coulomb lattice can yield to shear stresses.
\citet{td95} proposed the picture of ``starquakes'' --- sudden 
fractures and displacements of the crust, which shake the magnetosphere and trigger 
bursts. Cracks involving void formation are impossible in a neutron star crust 
because of the huge hydrostatic pressure \citep{jon03},
and slips are forbidden by the strong magnetic fields unless the slip plane is aligned 
with the magnetic flux surfaces \citep{ll12}. 
A plausible yielding mechanism is a plastic flow.
It is triggered when the elastic strain exceeds a critical value $\scr\simlt 0.1$,
where the maximum value of $\scr\sim 0.1$ describes the strength of an ideal crystal.
The lattice failure was  demonstrated on the microscopic level using 
molecular dynamic simulations \citep{hk09,ch10}.

A growing magnetic stress may be applied from the evolving core and 
reach the maximum elastic stress $\scr\mu$
where $\mu\sim 10^{30}$~erg~cm$^{-3}$ is the shear modulus of the lower crust;
then the crust will experience a shear flow.
Magnetic stresses can also be generated in the crust itself, due to Hall drift \citep{td96a,pp11}.
This leads to a thermoplastic instability \citep{bl14}, which 
launches thermoplastic waves, relieving the stress.
The propagating wave burns magnetic energy, resembling the deflagration front in combustion physics. Its speed is 
$v\sim (\chi B^2/4\pi \eta)^{1/2}$ where $\chi\sim 10$~cm$^2$~s$^{-1}$ is the heat 
diffusion coefficient and $\eta$ is the viscosity coefficient of the plastic flow.
\AB{Crustal flows are capable of releasing significant magnetic energy 
(e.g. \citealt{laa+15}) with a complicated temporal pattern.}
% They are likely to occur in avalanches sustained due to the excitation of short Hall waves in the avalanche \citep{llb16}.

\subsubsection{Observational signatures of internal instabilities}

Internal dynamics of magnetars shape their observational properties in three basic ways:
(1) glitches in the rotation rate, (2) internal heating and increased surface luminosity,
and (3) twisting of the external magnetosphere. This leads to 
rich phenomenology of magnetar activity described in \S2-4,
from giant flares to timing anomalies to persistent hard X-ray 
emission and hot spots on the magnetar surface.
Here we briefly discuss internal glitches; mechanisms of internal heating and 
magnetospheric phenomena are described in \S5.2-5.4.

A common internal mechanism for glitches in pulsars is related to neutron superfluidity
\citep{ai75}. The magnetospheric spin-down torque is applied to the crustal lattice 
and not directly to the free neutrons which dominate the star's moment of inertia. 
Neutrons in the lower crust are expected to form superfluid and their spin down
can lag behind, i.e. the neutrons rotate slightly faster than the crustal lattice. 
Vorticity of the neutron superfluid is quantized into vortex lines, and the lag 
is the consequence of the vortices being pinned to the lattice nuclei.
When the vortices become unpinned, they are allowed 
to move away from the rotation axis and the superfluid can spin down;
its angular momentum is passed to the lattice in this event, producing a spin-up glitch.
\citet{td96a} argued that sudden starquakes can cause such unpinning.
Alternatively, the superfluid vortices can be unpinned due to a 
heating episode \citep{le95}, and this is expected to occur in a 
thermoplastic wave.

%#####################################################

\subsection{Internal Heating and Surface Emission}
\label{sec:heating}

Heat that is initially stored in a nascent neutron star is eventually lost to neutrino emission 
and surface radiation. As a result, a kyr-old neutron star is normally expected to have an 
internal temperature $\Tc\sim 10^8$~K and a surface temperature $T_s\sim 10^6$~K 
\citep{yp04}, which corresponds to a surface luminosity 
$L_s\sim 10^{33}$~erg~s$^{-1}$.
In contrast, the surface luminosities of classical active magnetars are around 
$L_s\approx 10^{35}$~erg~s$^{-1}$.
For a neutron star of radius $R\approx 10-13$~km, this luminosity
corresponds to effective surface temperature 
$T_s=(L_s/4\pi R^2 \sigma_{\rm SB})^{1/4}\approx 4\times 10^6$~K, 
where $\sigma_{\rm SB}$ is the Stefan-Boltzmann constant. 
This indicates that magnetars are strongly heated by some mechanism.

\subsubsection{Heating of the core}

The high surface luminosity may be associated with a strong heat flux 
from the core, which implies an unusually high $\Tc$ \citep{td96a}. 
A steady heat flux to the surface is established on the conduction timescale 
of $1-10$~yr, which is much shorter than the magnetar age.
Then the steady heat diffusion determines the relation between $\Tc$ and $T_s$.
The main drop of temperature from $\Tc$ to $T_s$ occurs in the blanketing envelope
in the upper crust, in particular where $\rho<10^{9}$~g~cm$^{-3}$
\citep{yp04}. The $\Tc$-$T_s$ relation depends on the magnetic field $\bB$ in the blanket
and its chemical composition \citep{pyc+03}.
% Detailed calculations show that sustaining $L_s\sim 10^{35}$~erg~s$^{-1}$ 
% requires $\Tc\simgt 6\times 10^8$~K under most favorable assumptions \citep{bl16},
% in particular assuming a light-element blanket, which conducts heat better than iron does.
Increasing the core temperature above $\sim 10^9$~K does not increase $L_s$ 
because the heat flux is lost to neutrino emission on its way through the crust
\citep{ppp15}. 
Thus $L_s\sim 10^{35}$~erg~s$^{-1}$ is naturally obtained for any $\Tc\simgt 10^9$~K.

However, the ability of magnetars to sustain $\Tc\sim 10^9$~K for 1-10~kyr is 
questionable. At such high temperatures, huge neutrino losses are expected in the core itself
\citep{yp04}.
A minimum neutrino cooling rate is found in non-superfluid cores due to 
modified URCA reactions, $\dot{q}_\nu\sim 10^{21} T_9^8$~erg~cm$^{-3}$~s$^{-1}$.
% \citep{fm79}.
A possible transition to superfluidity at a critical temperature 
$\Tcrit\simgt 10^9$~K would only increase the cooling rate, as a 
result of continual breaking and formation of Cooper pairs \citep{frs76}.
% \citep{flo76,kgy+06,plps09}.
% \citep{flo76}.
In general, only a very powerful heat source can compete with neutrino losses at 
$\Tc\sim 10^9$~K.
A recent analysis by \citet{bl16} shows that heating by ambipolar diffusion may 
sustain the observed surface luminosity $L_s\sim 10^{35}$~erg~s$^{-1}$ if the internal 
magnetic field is ultrastrong, $B\simgt 10^{16}$~G, but only for a time shorter than 1~kyr.

\subsubsection{Heating of the crust}

An alternative scenario of internal heating assumes a cool core and a 
heat source in the crust. Then the radial temperature profile $T(r)$ peaks in the crust. 
Most of the heat is conducted to the core and lost to neutrino emission,
however a fraction reaches the stellar surface and could power the observed $L_s$.
Requirements of this scenario were investigated by 
% \citet{kyp+06,kpyc09,kkp+14}.
\citet{kkp+14}.
They placed a phenomenological heat source at various depths 
without specifying its mechanism, and 
found that sustaining the surface luminosity $L_s\approx 10^{35}$~erg~s$^{-1}$ 
requires a heating rate $\dot{q}_h>3\times 10^{19}$~erg~s$^{-1}$~cm$^{-3}$
at shallow depths $z\simlt 300$~m. 

Two possible mechanisms for converting magnetic energy to heat
are mechanical dissipation in the failing crust and ohmic heating.
Both obey strong upper limits \citep{bl16}. In particular, 
mechanical heating can only occur in the solid phase below the melted ocean,
and its rate cannot exceed $\dot{q}_{\max}\sim 0.1\mu\dot{s}$, where 
$\mu\sim 10^{28}\rho_{12}$~erg~cm$^{-3}$ is the shear modulus of the lattice and 
$\dot{s}$ is the strain rate of the deformation. This leads to an upper limit on
the persistent surface luminosity, which falls short of $10^{35}$~erg~s$^{-1}$.

The rate of ohmic heating $\dot{q}_h=j^2/\sigma$ is also limited, due to the 
high electric conductivity of the crustal material: $\sigma\sim 10^{22}$~s$^{-1}$ 
in the relevant range of temperatures and densities in the upper crust \citep{ppp15}. 
The timescale for dissipating electric currents that sustain variations $\delta B$
on a scale $\ell$ is 
$\tohm\sim 4\pi\sigma \ell^2/c^2\sim 4\times 10^4\, \sigma_{22}\,\ell_{\rm km}^2$~yr.
Ohmic dissipation could provide the required heating if two demanding conditions 
are satisfied:
(1) $\tohm$ is sufficiently short, comparable to the magnetar age of 1-10~kyr,
which requires a small scale of the field variations $\ell\sim 3\times 10^4$~cm, and
(2) the field variations on this scale must be huge, $\delta B \sim 10^{16}$~G, to provide 
$\dot{q}_h\sim (\delta B)^2/8\pi \tohm>3\times 10^{19}$~erg~s$^{-1}$~cm$^{-3}$.

It was proposed that ohmic and mechanical heating are boosted by Hall drift, which can 
transport magnetic energy to the shallow layers \citep{jon88} and develop 
large current densities \citep{gr92}.
%,vco00}.
Heating of the crust reduces its conductivity, leading to the coupled evolution of 
temperature and magnetic field. This magneto-thermal evolution was studied numerically
(e.g. \citealt{pmg09,vrp+13}), showing inhomogeneous 
heating of the stellar surface on timescales comparable to the magnetar ages.
The results suggest that the magneto-thermal evolution of the crust may explain 
the observed properties of a broader class of neutron stars, not only magnetars.

For simplicity, the simulations assumed that the crust is static, and this assumption 
needs to be relaxed in more realistic models.
Hall drift in the upper crust induces magnetic stresses exceeding 
the maximum elastic stress $\sigma_{\max}\sim 0.1\mu$, and the crust must flow, 
releiveing the stress. This leads to an upper limit on Hall-driven dissipation,
rendering it hardly capable of sustaining the surface luminosity of classical 
persistent magnetars \citep{bl16}. 
The full magneto-thermo-plastic evolution has been simulated in a one-dimensional 
model by \citet{llb16}. It showed intermittent heating 
through avalanches developing due to the excitation of short Hall waves in the avalanche,
which may explain the activity of transient magnetars.
% Future models will need to study this evolution in two and three dimensions.

\subsubsection{Transient heating and flare afterglow}

\citet{let02}
explored how a sudden deposition of heat in the crust could power the 
afterglow of the 1998 giant flare from SGR~1900+14. They found that heating 
would need to be approximately uniform throughout the upper crust, which implies 
enormous heat per unit mass in the surface layers $z< 100$~m.
It is, however, unclear how magnetic energy could be 
suddenly dissipated in the shallow layers,
which should behave as a liquid ideal conductor during the flare.
An additional complication is that the afterglow spectrum was nonthermal \citep{wkg+01}, 
suggesting magnetospheric origin. 
Nevertheless, the phenomenological picture of sudden crustal heating was 
applied to several transient magnetars, using detailed time-dependent 
simulations of heat conduction \citep[e.g.][]{rip+13,skc14}. 
The model was found capable of reproducing the observed light curves of 
some transient magnetars, except the cases where the emission area was 
observed to shrink with time. A mechanism explaining this shrinking will
be discussed in \S\ref{sec:untwist}.

A concrete mechanism for sudden crustal heating by a magnetospheric flare
was recently described by \citet{lb15}. 
They showed that Alfv\'en waves generated by the flare are quickly absorbed by the star 
and damped into plastic heat in the solid crust immediately below the melted ocean. 
A fraction of the deposited heat is eventually conducted to the stellar surface, 
contributing to the surface afterglow months to years after the flare.

\subsubsection{Surface spectrum}

The spectrum of thermal radiation emerging from the hot magnetar 
deviates from a Planck spectrum due to radiative transfer effects,
which are sensitive to the magnetic field \citep[see][for a review]{pdp15}.
An interesting potential spectral feature is the ion cyclotron line at energy
$\hbar ZeB/Am_pc\approx 0.63\,(Z/A)B_{14}$~keV.

The first detailed simulations of magnetar spectra assumed a fully ionized
hydrogen atmosphere \citep{hl01b,oze01,zts+01};
calculations have also been performed for a broader class of models.
Instead of forming a low-density atmosphere, the surface may
condense into a Coulomb liquid \citep{rud71,ml06} with a high density
$\rho\sim 10^5 AZ^{-3/5}B_{14}^{6/5}$~g~cm$^{-3}$, where $A$ and $Z$ are the 
ion mass and charge numbers. In this case, the emerging spectrum depends on the 
reflectivity of the liquid surface \citep{tzd04}.
It is also possible that the liquid is covered by a thin low-density atmosphere
\citep{spw09}.
% ,psv+12}.

Magnetar fields exceed the characteristic field 
$\BQ=m_e^2c^3/\hbar e\approx 4.4\times 10^{13}$~G 
defined by $\hbar \omega_B=m_ec^2$, where $\omega_B=eB/m_ec$. 
As a result, the photon normal modes are significantly 
impacted by the QED effect of vacuum polarization as well as plasma polarization.
Vacuum polarization dominates at densities 
$\rho\ll\rho_V\sim 1\, B_{14}^2 (\hbar\omega/1{\rm ~keV})^2$~g~cm$^{-3}$
and defines two linearly polarized normal modes:
the ordinary or O-mode (polarized in the ${\mathbf k}$-$\bB$ plane, 
where ${\mathbf k}$ is the photon wavevector) and the extraordinary or X-mode
(polarized perpendicular to the ${\mathbf k}$-$\bB$ plane).
At densities $\rho\gg \rho_V$ the plasma polarization dominates, 
which also gives O- and X-modes with highly elliptical polarizations. 
As photons pass through the layer with $\rho=\rho_V$
the polarization ellipse rotates by $90^{\rm o}$, effectively switching the modes
$O\leftrightarrow X$ \citep{lh03}, and the photon begins to see a different opacity. 
This effect tends to deplete the spectrum at high energies 
and also weakens the cyclotron absorption line.

% The surface is not homogeneous and 
\AB{In principle, the surface spectrum contains information on the surface magnetic field,
 however in practice extracting this information is extremely difficult.}
The cyclotron line is hard to detect when its position is smeared 
out due to variations in the surface magnetic field.
\AB{Besides $\bB$, the surface spectrum depends on the chemical composition 
and possible condensation, and disentangling all the effects is a challenging task.} 
% disentangling all the effects influencing the surface spectrum is difficult.
% In view of the uncertainties in the chemical composition and possible condensation, 
% disentangling all the effects influencing the surface spectrum is difficult.
\AB{Attempts to infer the surface $\bB$ from the observed 1-10~keV spectrum 
(e.g. \citealt{ggo11}) are further complicated by the superimposed 
magnetospheric emission.}
% The existing models suggest that the overall shape of the surface spectrum does 
% not dramatically deviate from a simple Planck spectrum. 

The inhomogeneity of temperature and magnetic field on the stellar surface
leads to flux pulsations with the magnetar spin period, 
which are influenced by the anisotropy of surface radiation and gravitational light 
bending. The surface radiation has a component beamed 
along the local $\bB$ and a broader fan component peaking perpendicular to $\bB$ \citep{zpsv95}. 
\AB{The observed pulse profiles are also strongly influenced by photon scattering in 
the magnetosphere.}

\subsubsection{Polarization}

Thermal radiation diffusing toward the magnetar surface is dominated by the X-mode 
photons, which have longer free paths and hence escape from deeper and hotter 
atmospheric layers. This radiation is linearly polarized with the electric field perpendicular
to the $\bk$-$\bB$ plane. Radiation emerging from a condensed surface is also 
linearly polarized; polarization of the radiation emitted by gaseous and condensed 
surfaces is discussed in detail by \citet{ttg+15}.

Polarization measured by a distant observer is influenced by two factors. 
(1) The observer receives surface radiation from regions with different magnetic fields,
which implies some averaging of the polarization.
(2) As the photons propagate through the magnetosphere, they ``adiabatically track'' 
the local normal mode defined by the local $\bB$: the polarization vector of the 
X-mode shifts so that it stays perpendicular to the $\bk$-$\bB$ plane.
This tracking ends and the polarization angle freezes when the photon reaches the
``adiabatic'' radius $r_{\rm ad}\sim 80 R\, B_{s,14}^{2/5}(\hbar\omega/1{\rm~keV})^{1/5}$,
where $B_s$ is the surface magnetic field \citep{hsl03}.
Therefore, effectively the observed polarization angle is controlled by the direction of 
$\bB$ at $r_{\rm ad}$ rather than at the stellar surface.

\subsection{Flares}

Magnetar flares and bursts emit hard X-rays and hence must be produced outside 
the neutron star. An ``internal trigger'' scenario assumes that the flares are
 triggered by an instability that leads to sudden ejection of magnetic 
energy from the core to the magnetosphere. The trigger 
could be an MHD instability in the liquid core or a sudden failure of the solid crust 
in response to a slow buildup of magnetic stresses at the crust-core boundary
\citep{td95,td01}. Excitation of core motions with displacement $\xi$  carries energy 
up to $\sim \Emag (\xi/R)^2$, where $\Emag\sim 10^{48}$~erg is the putative 
magnetic energy of the core. A displacement $\xi\sim R$ would strongly deform the 
magnetosphere and trigger a giant flare, however it implies a huge 
energy budget for the event, comparable to the entire $\Emag$.
A smaller $\xi$ could provide sufficient energy for a giant flare,
however  $\xi\ll R$ makes energy transfer to the magnetosphere inefficient \citep{lin14}.

An alternative ``external trigger'' scenario assumes a gradual deformation of the 
magnetosphere and the build up of the ``free'' magnetic energy followed by its 
sudden release, similar to solar flares \citep{lyu03,gh10,pbh13}.
This scenario requires only that the crustal motions pump an external twist 
faster than the magnetosphere damps it. This condition 
can be satisfied by thermoplastic crustal motions.
% Multiple  flares are expected to occur over the lifetime of a magnetar,
% as long as magnetic energy continues to be gradually 
% transferred from the star to the magnetosphere.

\subsubsection{Magnetospheric instability}

The magnetosphere twisted by surface shear motions becomes non-potential, 
$\nabla\times\bB\neq 0$, and threaded by electric currents $\bj=(c/4\pi)\,\nabla\times\bB$.
The magnetic energy still dominates over the plasma energy (including its rest mass),
and in the absence of strong ohmic dissipation the magnetosphere remains nearly 
force-free, $\bj\times\bB\approx 0$.
On a microscopic level, this condition corresponds to 
particles being kept in the ground Landau state in the strong magnetic field,
so that charges can only flow along $\bB$.

\begin{figure}
\includegraphics[width=5in,height=3in]{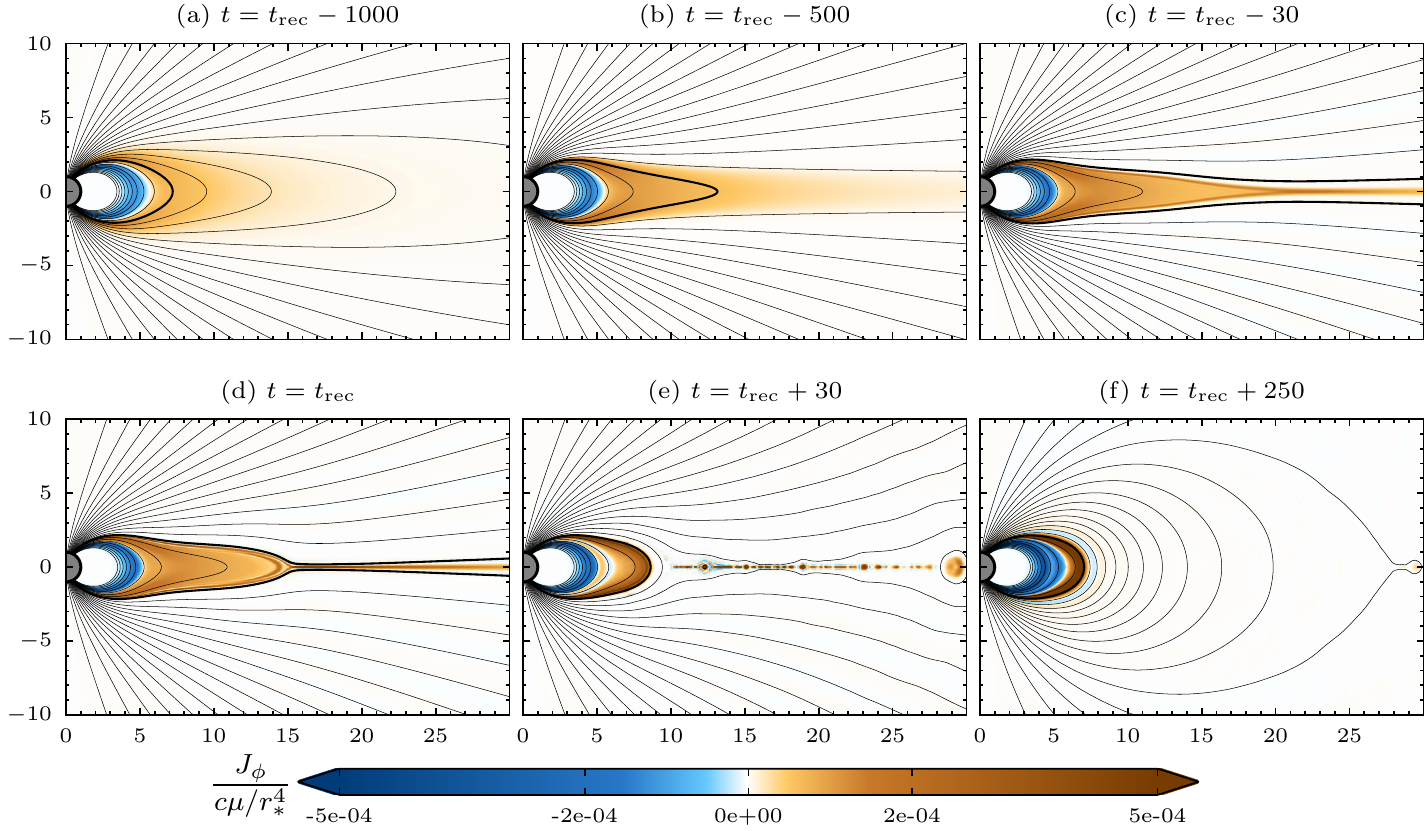}
\caption{Formation of the current sheet in an ``over-twisted'' magnetosphere.
Reconnection begins at $t_{\rm rec}$, panel (d). Color shows toroidal current density, 
lines are the poloidal magnetic field lines. One field line is highlighted in heavy black; 
it first opens and then closes again. Time is indicated in units of the light crossing 
time of the star. From \citet{pbh13}.
}
\label{recon}
\end{figure}

Strongly deformed magnetospheres are prone to global instabilities, 
as shown by the analysis of twisted magnetic equilibria (e.g. \citet{uzd02}).
An equilibrium magnetosphere obeys the 
equation $\nabla\times(\nabla\times\bB)=0$. The same equation describes the 
force-free solar corona, and
% (e.g. \citet{low78,aly84,wol95a}). 
the first models of static magnetar 
magnetospheres  \citep{tlk02} followed the self-similar solutions of \citet{wol95a}.
% ptzn09}.
More realistic configurations are obtained by numerical simulations
of the magnetosphere responding to surface shear \citep{pbh13}.
As the surface shearing continues, the twist angle $\ta$ grows until it reaches 
a critical value at which the magnetosphere becomes unstable.
Then it suddenly releases a significant fraction of the stored twist energy and 
produces a powerful flare. 
The flare involves a current sheet formation, its tearing instability and ejection of 
plasmoids, resembling coronal mass ejections from the sun \citep{ml94a}.
The flare development in axisymmetric geometry is shown in Figure~\ref{recon}.
Future simulations can significantly advance this picture
by relaxing axisymmetry and demonstrating the 
development of current sheets near the null points of more general magnetic configurations.

%%%%%%%%%%%%%%%%%%%%%%%%%%%%%
%%%%%%%%%%%%%%%%%%%%%%%%%%%%%

Current sheets spontaneously forming in over-twisted magnetospheres are fated 
to fast reconnection and energy dissipation. This process may naturally explain 
the fast onset of giant flares, as the dynamical timescale of the magnetosphere 
is $\delta t\sim R/c <1$~ms.
However, tension exists between the observed duration of the main peak,
$t_{\rm peak}\sim 0.3-0.5$~s, and the theoretical duration of the reconnection 
event $\sim 10^2R/c\sim 10^{-2}$~s \citep{uzd11}.

\subsubsection{Fast Dissipation and Radiation}

The main spike of a giant flare releases $\sim 10^{44}$-$10^{46}$~erg, 
which implies a huge energy density near the star. It is consistent with
dissipation of ultrastrong fields, releasing energy density  up to $B^2/8\pi\approx 4\times 10^{26}B_{14}^2$~erg~cm$^{-3}$.
The energy is immediately thermalized at a temperature $kT\simgt m_ec^2$,
creating a dense population of $e^\pm$ pairs and making the region 
highly collisional and opaque. In addition to sudden heating, reconnection 
generates strong Alfv\'en waves, which are ducted along the 
magnetic field lines and trapped in the closed magnetosphere. 
Part of the wave energy may cascade to small scales and dissipate
\citep{tb98}, and part is damped in the star.
% \citep{lb15}. 

The initial short spike of the giant flare must be emitted by a relativistic outflow
which traps radiation near the star, advects it and releases at a large radius 
$r>10^{10}$~cm \citep{td95}. This process resembles what happens in a 
cosmological gamma-ray burst \citep{mes02}.
A high Lorentz factor of the outflow $\Gamma$ helps the radiation to escape,
because the scattering 
optical depth is reduced as $\Gamma^{-3}$, and the 
photon-photon reaction $\gamma+\gamma\rightarrow e^++e^-$ is suppressed 
due to beaming of radiation within angle $\sim\Gamma^{-1}$.

After the initial spike, the temporarily opened field lines must close 
back to the star and trap a cloud of hot plasma (``fireball''), which produces the 
pulsating tail of the flare \citep{td95}. The fireball is confined if its energy $E_0$ is a 
fraction of $\RNS^3B^2/8\pi\sim 10^{45}B_{14}^2$~erg, so the field 
$B\simgt 10^{14}$~G is required by the observed energy radiated in the 
pulsating tail $E_0\sim 10^{44}$~erg. The trapped heat is 
gradually radiated away and the fireball gradually ``evaporates,'' layer by layer.
In the simplest model, the remaining heat $E(t)$ is proportional to the fireball volume 
while the luminosity $L=-dE/dt$ is proportional to the fireball surface area $A$. It implies 
$L\propto E^a$, where $a$ may be between $2/3$ (sphere) and 0 (slab), and 
gives  $L(t)=L_0(1-t/t_{\rm evap})^\chi$ with $0<\chi=a/(1-a)<2$ \citep{td01}.
This relation fits the observed envelope of the pulsating tail after $\sim 40$~s 
with  $t_{\rm evap}\approx 6$~min and large $\chi=3$, 
possibly indicating inhomogeneous structure of  the fireball  \citep{fhd+01}.
Further constraints on the model are provided by detailed calculations of neutrino 
losses \citep{goo11}.

The emission observed during the first 40~s of the flare 
(and after its main peak) requires an additional source, which
is brighter,  less modulated by rotation, and also harder. 
It was interpreted as emission from a heated corona surrounding the fireball \citep{td01}.
The emission observed after 40~s does not require additional heating and appears 
to be dominated by the evaporating fireball.
Its effective surface temperature 
$kT_s\sim 15\,L_{42}^{1/4}A_{13}^{-1/4}$~keV is smaller than 
the temperature inside the fireball by the factor of $\sim 10^{-2}$. 
The mean photon energy in a blackbody spectrum, $\hbar\omega\sim 3kT_s$, 
would be roughly consistent 
with the characteristic energy of observed hard X-rays, 
however the radiation spectrum is not Planckian. 
The spectrum must be shaped by radiative transfer in the fireball, and
a key feature of this transfer problem is the presence of two polarization 
states with drastically different free paths: the O- and X-mode.
Scattering of the X-mode is suppressed by the factor of 
$\sim (\omega/\omega_B)^2\sim 10^{-4}(\hbar\omega/10{\rm ~keV})^2 B_{14}^{-2}$.
Therefore, the X-mode photons dominate the energy transport and the emerging
luminosity; they escape from large Thomson optical depths where the O-mode 
photons are still trapped.
Since the E-mode cross section scales as $\omega^2$, photons of different
energies $\hbar\omega$ escape from different depths, leading to a flat 
spectrum at $\hbar\omega\simlt kT$ \citep{lub02}.
At high energies $\hbar\omega>kT$, the E-mode photons can  
split into O-mode photons in the ultrastrong magnetic field \citep{adl71}. 
The resulting theoretical spectrum of escaping radiation is far from Planckian
and resembles the observed broad spectrum.
\AB{The amplitude of  observed pulsations is likely affected by the scattering of 
fireball radiation in a continuing outflow from the star \citep{td01,vwb+16}.}

Similar fast dissipation of magnetic energy must occur in less powerful bursts,
which are much more frequent than the giant flares. Ordinary bursts with 
luminosity $L\ll 10^{42}$~erg~s$^{-1}$ may not be capable of producing 
fireballs with thermalized radiation, which explains their different spectra. 
The burst spectra are harder to model as they are more sensitive to the uncertainties 
in the dissipation mechanism, and detailed models have yet to be developed.

\subsubsection{Quasi-Periodic Oscillations}

The QPOs observed in giant flares (\S\ref{sec:qpos})
were interpreted as shear oscillations of the magnetar crust, which were studied by 
\citet{dun98}.
Simple QPO models assume that the crustal oscillations are decoupled
from the liquid core \citep{pir05,sa07}.
% \citealt{pir05,wr07,lee07,sks07}.
More realistic models allow for the crust coupling by the magnetic field
to the continuum of Alfv\'en waves in the liquid core
% and the core are coupled by the magnetic field lines
%  \citep{lev06,lev07,gsa06,csf09,ck11,vl11,vl12,gcf+13,gcs+14,pl14}.
\citep{lev07}.
%,gsa06}.
This creates significant uncertainties in the frequency spectrum of the global modes, 
as it depends on the magnetic configuration in the star. However some general 
features have been identified. 
The strongest oscillations are expected near the edges of the spectrum \citep{vl11}. 
The oscillations experience phase mixing and QPOs can be transient and delayed.
The models may be reconciled with the observed range of QPO frequencies
if neutrons are decoupled from the oscillations \citep{vl11,gcs+13,pl14},
which is possible if the neutrons are superfluid. 
High-frequency oscillations, in particular the 625-Hz QPO in the 1998 giant flare,
are most challenging to model. 

The magnetosphere attached to the oscillating crust is periodically
deformed \citep{gcs+14}, which could lead to periodic changes in its luminosity. 
The models predict small amplitudes of surface oscillations, much smaller 
than the stellar radius, and it remains unclear how they generate the 
observed 10-20 percent modulation of the X-ray luminosity.

% \citep{tel08,dw12,gcs+14}. 

\subsection{Gradual Energy Release in the Twisted Magnetosphere}
\label{sec:twist}

The magnetosphere can remain twisted for years between the flares,
explaining the magnetar activity described in \S3.
The presence of a magnetic twist $\nabla\times\bB\neq 0$ implies long-lived electric 
currents, which are accompanied by some ohmic dissipation. The 
magnetosphere always tends to slowly untwist,
and the released magnetic energy feeds its long-lived emission 
\citep{tlk02,bel09}.

\subsubsection{Electric discharge}

Magnetospheric electric currents can only flow along the magnetic field lines and their
ohmic dissipation occurs due to a small electric field 
$\Epar$ parallel to $\bB$. The electric field has three functions: 
(1) it maintains the electric current $\bj\parallel \bB$ demanded by $\nabla\times\bB\neq 0$, 
(2) it regulates the dissipation rate $\Epar j$ and the observed nonthermal 
luminosity, and 
(3) it determines the evolution of $\bB$ in the untwisting magnetosphere
according to the Maxwell-Faraday equation $\partial\bB/\partial t=-c\,\nabla\times\bE$.

The longitudinal voltage between the two footprints `1' and `2' of a magnetospheric 
field line, $\Phi_\parallel=-\int_1^2 E_\parallel dl$, controls 
the ohmically released power $L \approx I \Phie$, where 
$I<I_{\max}\sim c\mu/R^2$ is the net  electric current circulating through the 
% twisted 
magnetosphere and $\mu$ is the magnetic dipole moment of the star.
The observed nonthermal luminosities of magnetars 
typically require $\Phie\sim 10^{10}$~V. The voltage 
was proposed to be regulated 
% through continual
by 
$e^\pm$ discharge: 
when $\Phie$ exceeds a threshold value, an exponential runaway of $e^\pm$ 
creation occurs until the pair plasma screens $E_\parallel$  \citep{bt07}. 
As the plasma leaves the discharge region, $E_\parallel$ grows again and the discharge 
repeats, resembling continual lightning. 

In a magnetar magnetosphere, the discharge is triggered when an electron accelerated 
to Lorentz factor $\gamma$ begins to resonantly scatter X-rays streaming from the star.
The X-ray photon is blueshifted  in the electron rest frame by the factor of $\sim\gamma$,
and resonant scattering occurs when the blueshifted photon frequency 
matches the cyclotron frequency $\omega_B=eB/m_ec$.
In the ultrastrong field near the magnetar this typically requires $\gamma\simgt 10^3$. 
The scattered photon has a high energy 
$\sim \gamma^2$~keV and quickly converts to $e^\pm$ in the strong 
magnetic field. This process gives the threshold voltage $\Phie\sim 10^9-10^{10}$~V.
The discharge can be modeled ab initio using "particle in cell" (PIC) method that 
follows plasma particles individually in their collective electromagnetic field. First 
axisymmetric PIC simulations of twisted magnetospheres have been performed 
recently \citep{cb16}.

\subsubsection{Resistive untwisting and shrinking hot spots}
\label{sec:untwist}

The resistive evolution of twisted magnetospheres has only been studied thus far 
in axisymmetric geometry.
% \citep{bel09}. 
In this case the magnetosphere may be 
described as a foliation of axisymmetric flux surfaces labeled by magnetic flux 
function $f$, and the twist of a magnetospheric field line $\psi(f)$ 
is the difference between the azimuthal angles of its footpoints on the star,
$\ta=\phi_2-\phi_1$.
The twist evolution is governed by the electrodynamic equation obtained from 
$\partial\bB/\partial t =-c\,\nabla\times \bE$,
\be
\label{eq:evol}
   \frac{\partial\psi}{\partial t}=2\pi c\frac{\partial\Phie}{\partial f}
     +\omega(f,t),
\ee
where $\omega=\dot{\phi}_2-\dot{\phi}_1$ is the applied shear rate at the stellar surface
\citep{bel09}.
When the fast crustal motions stop, so that $\omega\ll 2\pi c\, \partial\Phie/\partial f$, 
the untwisting phase begins.

Two distinct regions evolve in an untwisting magnetosphere: 
a ``cavity'' with $j=0$ and a ``j-bundle'' where the currents flow.
% \citep{bel09}.
The cavity is comprised of field lines that close near the star,
and has a sharp boundary along a flux surface $\ff$.
The process of resistive untwisting is the slow expansion of the boundary $f_\star$, 
which erases the electric currents in the j-bundle on a timescale
$t_{\rm ohm}\simlt\mu/cR\Phie$. 
As a result, the magnetospheric currents have the longest lifetime on magnetic 
field lines with large apex radii $R_{\max}\gg R$, i.e. the 
magnetar activity tends to be confined to field lines extending far from the star.

This electrodynamics implies a special observational feature: 
a  slowly shrinking hot spot on the magnetar surface.
The footprint of the j-bundle is expected to be hotter than the rest of the stellar surface,
because it is bombarded by relativistic particles from the $e^\pm$ discharge.  
As the j-bundle slowly shrinks so does its footprint. 
The theoretically expected relation between the spot area $A$ and luminosity $\Lum$
is given by $\Lum\sim 1.3 \times 10^{33} K\,A_{11}^2$~erg~s$^{-1}$ where 
$K=B_{14} \Phi_{\parallel 9} \psi$.
% \citep{bel09}. 
Such shrinking hot spots have been 
observed in seven transient magnetars (see \cite{bl16} for a compilation of data). 
The spot area and luminosity
decrease with time, and the observed slope of the $A$-$\Lum$ relation 
(controlled by the behavior of $\Phi_\parallel$) varies between 1 and 2. 
The typical timescale of this evolution, months to years, 
is also consistent with theoretical expectations, however there are outliers that 
require a more detailed modeling.

Transient magnetars also show different behavior, which is inconsistent with 
the simple model of untwisting  magnetosphere attached to a static crust.
For instance, 1E~1547$-$5407 displays complicated activity \citep{khd+12},
possibly due to repeated twist injection, which prevents
a clean, long twist decay. In addition, a different emission component
may  result from cooling of a suddenly heated crust, as discussed in 
Section~\ref{sec:heating}.

\subsubsection{Nonthermal X-rays}
\label{sec:nonth}

The dissipated twist energy is given to the $e^\pm$ plasma in the j-bundle, 
which can radiate it away in two ways: the particles can hit the stellar surface or 
pass their energy to the ambient X-rays streaming from the star, mainly through  
resonant Compton scattering \citep{tlk02,bh07}. 

Scattering of a keV photon by a particle with Lorentz factor $\gamma$ 
boosts the photon energy by a factor of $\sim\gamma^2$. Early work \citep{tlk02}
proposed that the magnetosphere is filled with mildly relativistic electrons 
which scatter the thermal surface radiation and produce 
a soft power-law tail in the X-ray spectrum.
The resonance scattering condition implies that mildly relativistic Comptonization 
can occur where $B\sim 10^{11}$~G which corresponds to $\hbar eB/m_ec\sim 1$~keV;
this region is typically at radii $r\sim 10\RNS$. Detailed Comptonization models 
with ad hoc particle distributions were developed 
\citep{lg06,ft07,ntz08}
and found capable of reproducing the observed 1-10~keV spectra.
A simplified resonant Comptonization model was implemented under XSPEC and 
used to fit the X-ray spectra of magnetars below 10~keV 
\citep{rzt+08,zrtn09}.

%%%%%%%%%%%%%%%%%%%%%%%%%%%%%
\begin{figure}
\parbox{6cm}{
\includegraphics[width=6cm]{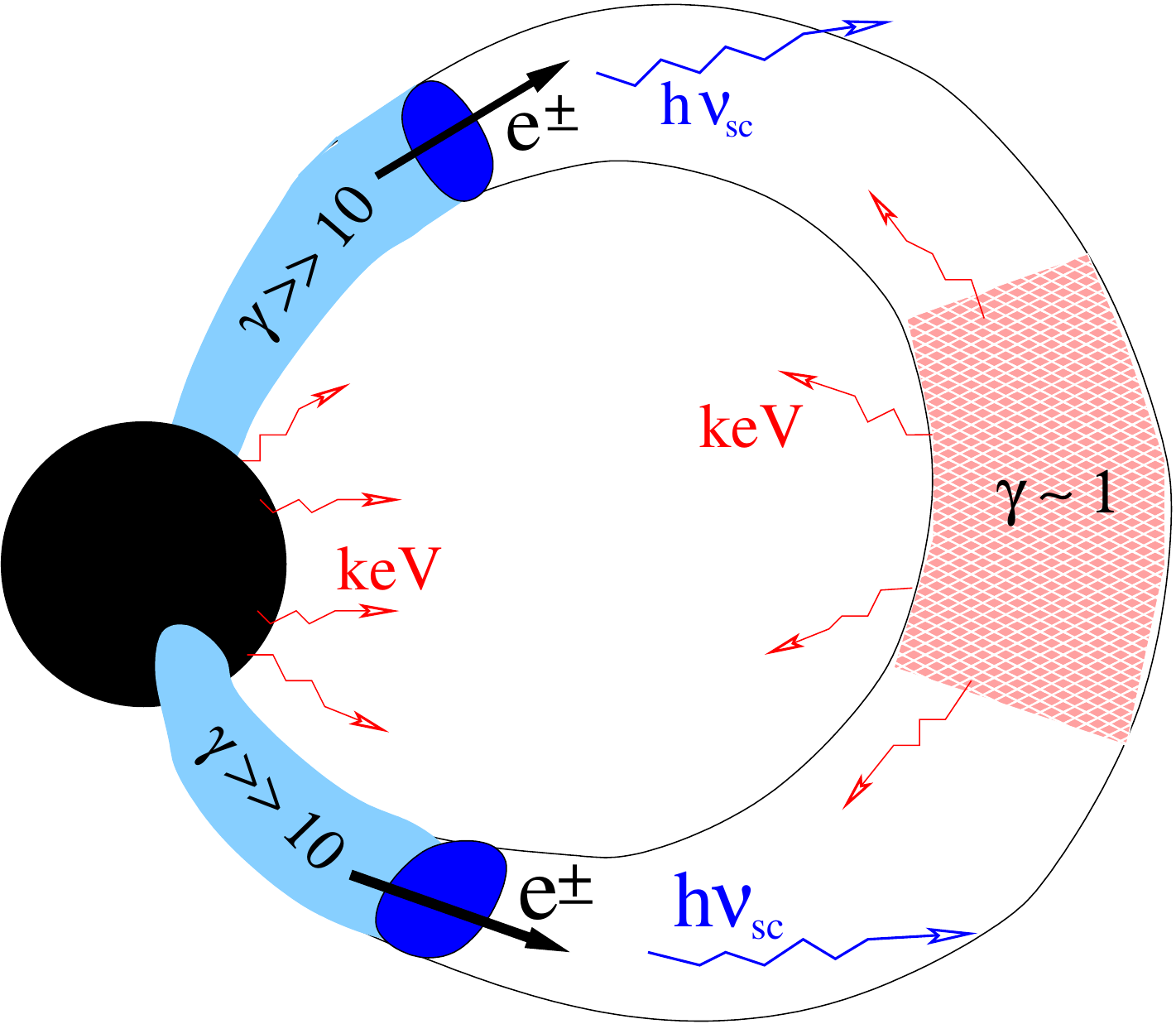}
% \caption{First.}
% \label{fig:2figsA}
}
\qquad
\begin{minipage}{6cm}
\includegraphics[width=6cm]{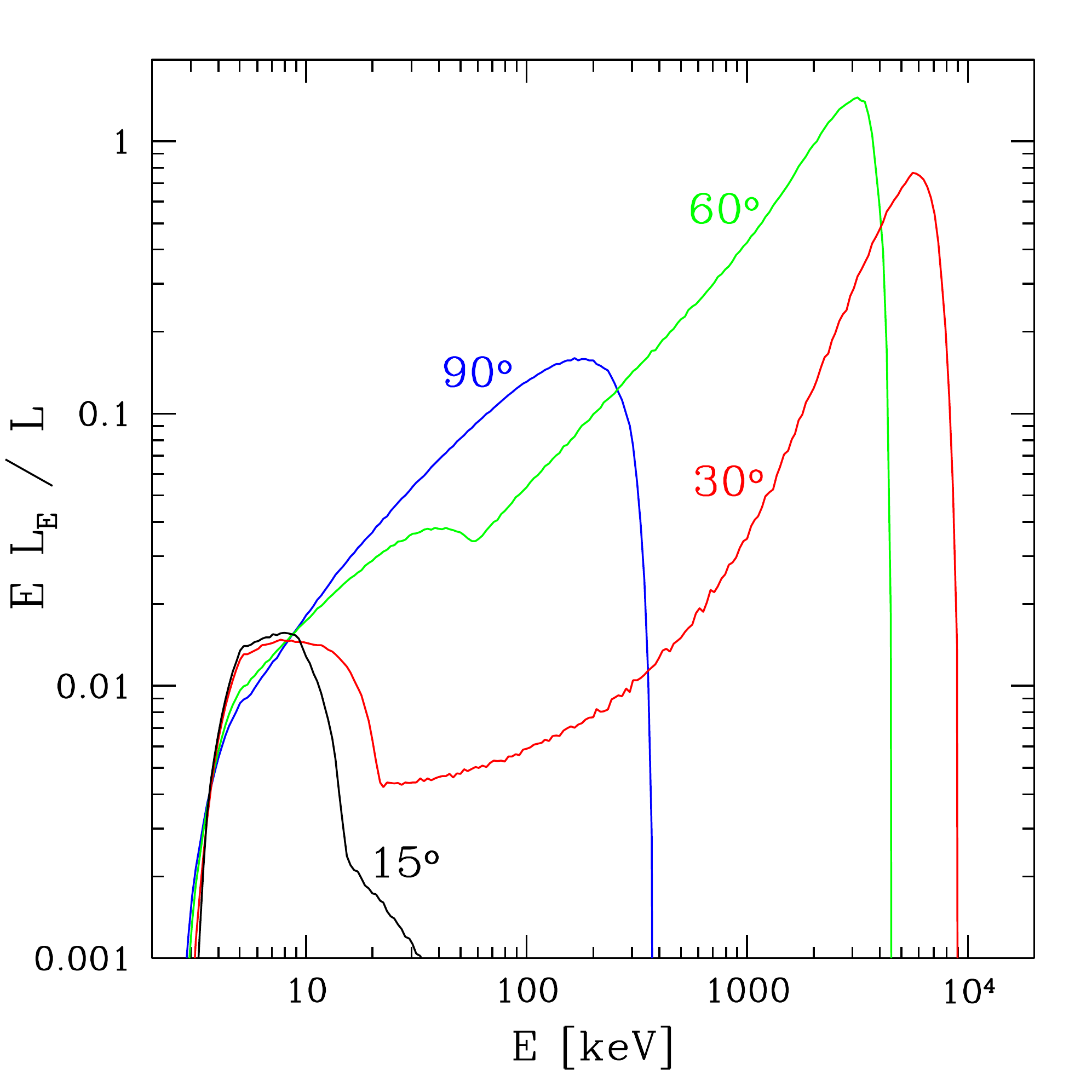}
% \caption{Second.}
% \label{fig:2figsB}
\end{minipage}
\caption{Left: A magnetic loop in the j-bundle. Relativistic particles are assumed to be
injected near the star (black sphere), and a large $e^\pm$ multiplicity $\M\sim 100$ 
develops in the `adiabatic' zone $B > 10^{13}$~G (shaded in blue). The outer part of 
the loop is in the radiative zone; here the resonantly scattered photons of energy 
$h\nu_{\rm sc}\sim 1\,B_{13}^2$~MeV escape and form the observed hard X-ray spectrum.
% shown in 
% Figure~\ref{fits}.
The outflow decelerates and eventually annihilates at the top of the loop (shaded in pink); 
here it becomes opaque to the thermal keV photons flowing from the star. 
Photons reflected from the pink region have the best chance to be scattered by 
the relativistic outflow in the lower parts of the loop, and control its deceleration.
The footprint of the j-bundle is heated by the relativistic backflow from the discharge region; 
in an axisymmetric model it forms a hot ring on the stellar surface.
Right: Radiation spectrum from the j-bundle viewed at four different 
angles with respect to the magnetic dipole axis. From \citet{bel13b}.
}
\label{fig:illustr}
\end{figure}
%%%%%%%%%%%%%%%%%%%%%%%%%%%%%

Recent work attempted a self-consistent calculation of the particle motion in the 
j-bundle \citep{bel13a,bel13b}, which led to the picture shown in 
Figure~\ref{fig:illustr}.
Pairs must be created with high Lorentz factors $\gamma$ and their subsequent motion 
is influenced by radiative losses due to resonant scattering. The upscattered
photons have high energies and convert to $e^\pm$ pairs in the region of $B>10^{13}$~G.
As a result, energy radiated near the star is processed into an outflow (``fountain'') of 
copious pairs with a standard profile of Lorentz factor $\gamma\approx 100(B/\BQ)$.
The outflow radiates away its kinetic energy in the outer zone $B\simlt 10^{13}$~G and 
comes to the top of the magnetic loop with $\gamma\sim 1$. 
Mildly relativistic Comptonization may occur in this outer region and 
influence the X-ray spectrum below 10~keV. However, most of the nonthermal 
luminosity is radiated above 10~keV in the lower parts of the loop, where the outflow 
has $\gamma\gg 1$. 

Regardless of the details of the electric discharge near the star, the  
$e^\pm$ fountain emits a power-law spectrum
$dL/d\ln E\propto E^{1/2}$.
It naturally produces a distinct hard component in magnetar spectra, which 
peaks and cuts off in the MeV band. However, the $E^{1/2}$ power law is predicted
only for spectra averaged over viewing angles.
The emission is beamed along the magnetic field lines within angle 
$\sim \gamma^{-1}\sim 0.1 B_{13}^{-1}$, 
and the overall angular distribution is determined by the field-line curvature.
The field may be approximated as dipolar
in the radiative zone $B\simlt 10^{13}$~G, which is relatively far from star.
In this approximation, the predicted spectrum varies with the viewing angle 
as shown in Figure~\ref{fig:illustr}.

This model provided good fits to the phase-resolved 
hard X-ray spectra of magnetars 
% \citep{ahk+13,aah+15,hbd14,vhk+14,thy+15}.
\citep{hbd14,aah+15,vhk+14}.
The main unknown geometric parameters are 
the angle between the magnetic dipole axis and the rotation axis, 
$\alpha_{\rm mag}$, and the angle between the line of sight and the rotation axis, 
$\beta_{\rm obs}$.
Remarkably, only a small region in the parameter space provided a good fit, 
which allows one to estimate $\alpha_{\rm mag}$ and $\beta_{\rm obs}$. The results suggest
that the magnetic dipole axis in magnetars is slightly misaligned with the rotation axis.  

The polarization of radiation upscattered in the magnetosphere was discussed
by \cite{fd11} and \citet{bel13b}. 
After resonant scattering the photon ``forgets'' its initial polarization; it becomes X-mode
with 75\% probability and O-mode 25\% probability. The escaping radiation
should be dominated by the X-mode, except at energies approaching 1~MeV, where 
photon splitting occurs $X\rightarrow O+O$  \citep{adl71}.
One may observe the O-mode polarization if the daughter photons from splitting do 
not scatter on the way out from the magnetosphere.
Future measurements of X-ray polarization can provide powerful diagnostic 
tools, taking the advantage of magnetar rotation (which gives a periodically changing viewing angle) and combining with the phase-resolved spectra.

A few other ideas were proposed for the origin of the hot plasma capable of emitting hard 
X-rays. \citet{hh05} discussed the possibility of shock formation by 
waves traveling in the magnetosphere. \citet{tb05} 
% and \citet{bt07}
proposed that a dense transition layer with $kT\sim 100$~keV  forms between 
the magnetosphere and the stellar surface and radiates bremsstrahlung photons.
This model does not, however, explain the variations of the hard X-ray spectrum with the 
rotational phase and why the emission at different energies peaks at different phases.

\subsubsection{Low-frequency Emission}

In ordinary radio pulsars, radio emission is believed to come from the open field lines 
that connect the star to its light cylinder --- this part of the magnetosphere is 
persistently active (twisted) and carries electric current 
$I_{\rm open}\sim \mu\Omega^2/2c$.
The maximum voltage induced by stellar rotation on the open field 
lines is $\Phipc\approx \mu \Omega^2/c^2$ \citep{rs75}. 
For typical magnetar parameters, $\mu\simgt 10^{32}$~G~cm$^3$ and 
$\Omega\sim 1$~rad$^{-1}$, one finds $e\Phipc/m_ec^2\sim 10^8$, more than
sufficient to sustain pair creation.

However, magnetar radio emission from the pair plasma in the open bundle may be 
undetectable for two reasons. (1) Pair discharge can limit the voltage to a much lower
value $\Phie\ll\Phipc$;  then the power dissipated in the open bundle is 
small: $\Phie I_{\rm open}\sim 10^{29} (\Phie/10^{10}{\rm V})$~erg~s$^{-1}$. 
With a reasonable radiative efficiency this implies a low radio luminosity,
well below $L_{\rm radio}\sim 10^{30}$~erg~s$^{-1}$ observed in \XTE~\citep{crh+06}.
(2) The radio beam from the open bundle
may be narrow and miss our line of sight.

The fact that radio pulsations are associated with outbursts (\S\ref{sec:multitemporal}) 
suggests a connection with a magnetospheric twist.
A scenario with strong dissipation on open field lines, $\Phie I_{\rm open}$, enhanced 
by the twist in the closed magnetosphere \citep{tho08b}, is problematic
--- this corresponds to $\partial\Phie/\partial f<0$ in \Eq~(\ref{eq:evol}) and the twist 
around the open bundle should be quickly erased. 
Radio emission can be produced by the closed j-bundle itself. It 
is much thicker and more energetic than the open bundle, and therefore capable of 
producing much brighter radio emission with a broad pulse. 
The untwisting magnetosphere in XTE J1810-197 had a j-bundle with 
magnetic flux $f_j\sim 3\times 10^2 \fopen$, and carried electric current 
$I\sim 10^5I_{\rm open}$. Then a reasonable efficiency of radio emission 
$L_{\rm radio}/\Phie I\sim 10^{-3}$ is sufficient to explain the observed 
radio luminosity \citep{bel09}.
Given the high plasma density in the j-bundle, 
$n\sim 10^{18}-10^{20}$~cm$^{-3}$, the plasma frequency in the $e^\pm$ outflow
may approach the infrared or even optical band. 
Thus, plasma processes in the j-bundle might contribute to emission from radio to optical
\citep{egl02}.

\subsection{Spin-down Torque}

The twisted magnetosphere of an active magnetar is somewhat inflated due to the 
additional pressure of the toroidal field $B_\phi^2/8\pi$. 
This inflation increases the open magnetic 
flux $f_{\rm open}$ that connects the star to its light cylinder and therefore increases 
the spindown torque applied to the star \citep{tlk02}. 
% Electric 
% currents flowing in an axisymmetric j-bundle with magnetic flux $f_j$ and a small 
% twist angle $\psi<1$  increase the effective magnetic moment of the star by 
% $\Delta\mu/\mu\sim (\psi^2/4\pi)\ln (f_j/f_{\rm open})$ \citep{bel09}. 
A strong increase in spindown is expected for strong twists 
$\psi>1$, which also produce a higher magnetospheric luminosity.

However, no strict general relation between the X-ray emission and spindown 
is expected, in particular in non-axisymmetric magnetospheres. 
The torque is sensitive to the 
behavior of the small flux bundle $f_{\rm open}$, a tiny fraction of the total magnetic flux 
of the star, $f_{\rm open}/f_{\rm total}\sim \RNS/\RLC\sim 10^{-4}$. 
The open flux is much smaller than $f_j$ and may be outside the j-bundle. 
Therefore, the spindown torque may react to changes 
in the magnetospheric twist with a delay or sometimes even anti-correlate with 
the X-ray emission, depending on the details of the magnetospheric configuration.

It was also proposed that the persistent high spindown rate is due to strong plasma 
loading of the magnetar wind \citep{hck99,txs+13}. This proposal posits that the true 
dipole component is much smaller than inferred from the standard spindown formula 
$B\approx 3\times 10^{19}(P\dot{P})^{1/2}$~G, and the magnetic energy required to 
feed the magnetar activity is stored in much stronger multipoles. A challenge for 
this scenario is that it needs a dense plasma outflow that would energetically 
dominate at the light cylinder and ``comb out'' the magnetic field lines,  increasing 
the open magnetic flux and boosting the spindown rate. 
It is unclear if e.g. seismic activity of the star would be able to drive such an outflow 
 \citep{tdw+00}.

Magnetospheric flares are expected to impact the magnetar spindown.
The existing flare simulations show that the magnetic flux connecting 
the star to its light cylinder $f_{\rm open}$ is dramatically increased during the flare, 
a result of strong inflation of the twisted field lines \citep{pbh13}. 
The spindown torque exerted on the star $\dot{J}\sim-f_{\rm open}^2/2\pi c P$ 
becomes enormous for a short time $\Delta t$ comparable to $\Omega^{-1}=P/2\pi$ in 
the simulation, and produces a sudden increase of the 
rotation period $\Delta P>0$ --- an ``anti-glitch''.
When applied to the 1998 August giant flare in SGR~1900+14, the model gives
$\Delta P/P\sim 10^{-4}$, consistent with the observed anti-glitch \citep{wkv+99}.
The predicted $\Delta P$ may vary in more realistic models 
where the flare is not axisymmetric and the light cylinder is far outside of the main 
flare region. This might explain the non-detection of an anti-glitch 
($\Delta P/P<5\times 10^{-6}$) in the exceptionally powerful giant flare of SGR~1806-20 
on December~24, 2004, despite 
the observed ejection of a powerful outflow during the initial spike of the flare. 
Note also that the above picture does not work for anti-glitches that are not associated 
with giant flares, such as the one reported by \citet{akn+13}.

%\bibliography{review.bib}

%\end{document}
%
%##################################################################
%##################################################################
%##################################################################

\section{CONCLUSIONS AND FUTURE WORK}
\label{sec:conclusions}

The magnetar model has now been used to predict, naturally and uniquely, 
a wide variety of remarkable phenomena and behaviors in sources that 
once seemed highly `anomalous.'
The now seemless chain of phenomenology from otherwise conventional radio pulsars through
sources previously known for radically different behavior makes clear that these objects are
one continuous family, with activity correlated with spin-inferred
magnetic field strength.  Recent advances in
the physics of these objects, from the core through the crust and
to the outer magnetosphere, hold significant promise.
Below are issues we believe hold potential for important progress
in the field in the near future, as well as remaining unsolved problems that are
worthy of more thought.

\begin{issues}[FUTURE ISSUES]
\begin{enumerate}

\item Why the `transient' magnetars are orders-of-magnitude fainter and significantly softer in X-rays
than the persistent sources remains an important puzzle.  
Continued X-ray and multi-wavelength follow-up of newly discovered magnetars, found using
all-sky X-ray monitors sensitive to outbursts, will
flesh out spin-property distributions, better constrain the Galactic
population of transient magnetars, and their outburst rates.

\item Future X-ray polarimetric observations of magnetars will
test basic predictions of quantum electrodynamics, and
illuminate the geometry of the magnetic field for comparison with that
inferred from modeling of phase-resolved hard X-ray spectra.

\item Monitoring of high-magnetic-field radio pulsars, particularly their behavior near
glitch epochs and at rare times of magnetar-type activity, will help clarify the 
onset of instabilities with increasing spin-inferred magnetic field.

\item Detailed studies of magnetar wind nebulae and associations with TeV
emission may be useful for calorimetric
determinations of past magnetar activity.

\item Numerical simulations 
of ambipolar diffusion in the core and advanced magneto-thermoplastic models of crust evolution may shed light on how magnetars become active, and permit quantitative predictions for their transient and persistent activity.

\item First-principle simulations of twisted magnetospheres have now become possible using the plasma particle-in-cell
method. This technique has recently been successfully applied to ordinary radio pulsars and can be applied to magnetars.

\item Three-dimensional simulations of relativistic reconnection in the magnetosphere can give more realistic models of bursts and giant flares.

\item The results of recent modeling of internal heating and post-flare QPOs provide strong constraints on magnetar interior and can be used to infer basic properties such as superfluidity in the core and the strength of the internal magnetic field.

\item 
Modelling of the remarkable glitches and anti-glitches observed in magnetars and their 
radiative signatures may become possible in the near future as part of detailed 
simulations of magnetar interiors that include superfluid neutrons.

\end{enumerate}
\end{issues}

\section{ACKNOWLEDGEMENTS}
\label{sec:ack}

VMK acknowledges funding from the Natural Sciences \& Engineering Research Council
of Canada, the Canada Research Chairs program, the Canadian Institute for
Advanced Research, the Lorne Trottier Chair in Astrophysics \& Cosmology,
and Les Fonds de recherche du Qu\'ebec -- Nature et technologies.
AMB acknowledges NASA grant NNX13AI34G and a grant from the Simons Foundation 
(\#446228, Andrei Beloborodov)

%\bibliography{journals,psrrefs,modrefs,arrefs,review}
%\bibliographystyle{ar-style2}

\end{document}